\documentclass[epj]{svjour}
\usepackage{graphics}
\def\be {\begin{equation}}
\def\ee {\end{equation}}
\def\nn {\nonumber}
\def\bea {\begin{eqnarray}}
\def\eea {\end{eqnarray}}

\newcommand{\om}{\omega}  
\newcommand{\vk}{\vec k}

\newcommand{\vp}{\vec p}
\newcommand{\vV}{\mathbf{V}}
\newcommand{\vn}{\vec {\nabla}}

\newcommand{\bp}{\vp}
\newcommand{\del}{\partial}

\begin{document}
\title{Shear viscosity and electrical conductivity of relativistic fluid in 
presence of magnetic field: a massless case}
\author{Jayanta Dey\inst{1} \and Sarthak Satapathy\inst{1} \and Prasanta Murmu\inst{2,1} \and Sabyasachi Ghosh\inst{1}} 
%
\institute{Indian Institute of Technology Bhilai, GEC Campus, Sejbahar, Raipur 492015, Chhattisgarh, India 
\and Indian School of Mines Dhanbad 826004, Jharkhand, India.}
\date{Received: date / Revised version: date}
%
\abstract{
We have explored the shear viscosity and electrical conductivity calculations for 
bosonic and fermionic medium, which goes from without to with magnetic field picture and then their simplified massless expressions. In presence of magnetic field, 5
independent velocity gradient tensors can be designed, so their corresponding proportional coefficients, connected with the viscous stress tensor provide us 5
shear viscosity coefficients. In existing litterateurs,
two sets of tensors are available. Starting from them, present work has obtained two sets of expressions for 5 shear viscosity coefficients, which can be ultimately classified into three basic components - parallel, perpendicular and Hall components as one get same for electrical conductivity at finite magnetic field.
Our calculations are based on kinetic theory approach in relaxation time approximation.
Repeating same mathematical steps for finite magnetic field picture, which traditionally practiced for without field case, we have obtained 2 sets of 5 shear viscosity components, whose final expressions are in well agreements with earlier references, although a difference in methodology or steps can be clearly noticed.
Realizing the massless results of viscosity and conductivity for Maxwell-Boltzmann,
Fermi-Dirac and Bose-Einstein distribution function, we
have applied them for massless quark gluon plasma and hadronic matter phases, which can provide us a rough order of strength, within which actual results will vary during quark-hadron phase transition. Present work also indicates that magnetic field might have some role for building perfect fluid nature in RHIC or LHC matter. The lower bound expectation of shear viscosity to entropy density ratio is also discussed.
%
} 
\authorrunning{Dey, Satapathy, Murmu, Ghosh}
\titlerunning{Viscosity and conductivity in presence of magnetic field}
\maketitle

\section{Introduction}
Heavy-ion collisions have been the subject of intensive research to 
extract information about the nuclear properties of matter in extreme 
conditions like high temperature and high external magnetic fields. 
This field covers many interesting phenomena such as magnetic catalysis~\cite{Shovkovy}, 
chiral magnetic effect~\cite{Bzdak:2011yy}\cite{Elia}, inverse magnetic catalysis
~\cite{Bali1,Bali2,Mueller} etc. A detail discussion on the effects of magnetic field 
in quantum field theory has been addressed in Ref.~\cite{Miransky}.
A verification of these anticipated results is possible by studying QCD matter under 
the influence of electromagnetic fields. A review of the phase structure of QCD in the presence of magnetic 
field has been given in Ref.~\cite{Andersen}.

Ref.~\cite{Tuchin:2013ie} shows analytically that fields produced in RHIC and 
LHC can reach up to $m_{\pi}^2$ and $10m_{\pi}^2$ respectively after collision. 
Possible space time evolution of electromagnetic fields, generated in heavy-ion collisions
are well discussed in Refs.~\cite{Deng:2012pc,Satow:2014lia,Skokov:Illarionov}. 
Applying that space time evolution of magnetic field information to the hydrodynamical 
expansion of quark-gluon plasma will construct a detail expansion dynamics, which is
commonly known as magneto hydrodynamic (MHD). Under the influence of a strong
magnetic field the properties of quark-gluon plasma in heavy-ion collisions have been 
studied under the generalized framework of Bjorken flow ~\cite{Roy:2015kma,Pu:2016ayh}.
In the limit of ideal magnetohydrodynamics, i.e., for infinite conductivity, and irrespective 
of the strength of the initial magnetization, the decay of the fluid energy density  with proper 
time is the same as for the  “Bjorken flow” without magnetic field. It has been 
found in Ref.~\cite{Pu:2016ayh} that under the influence of magnetic field the 
energy density and temperature decay slowly because of magnetic field. 
But taking into consideration the magnetic field produced in heavy-ion collisions 
the decay is suppressed. Numerical approaches to magnetohydrodynamics and it's 
application to heavy-ion collisions have been studied~\cite{Hongo:2013cqa,Inghirami:2016iru}. 
Impact of external magnetic field through
transport simulation can be noticed in Ref.~\cite{Das:2016cwd}.
For dissipative picture of MHD or transport simulation, field dependent transport coefficients might be
required as inputs and therefore, a parallel microscopic
calculation of transport coefficients in present of magnetic field~\cite{Landau} is an important
topic in heavy ion Physics community.
In recent time transport coefficients in presence of magnetic field are investigated in
Refs.~\cite{Tuchin,Li_shear,Asutosh,G_NJL_B,Nam,Tawfik,JD2,HRGB,HM3,Denicol,BAMPS,Sedrakian_el,Kerbikov:2014ofa,Nam:2012sg,Hattori:2016lqx,Manu1,Manu2,Feng_cond,Fukushima_cond,Arpan1,Arpan2,NJLB_el,Lata,BC,Manu_SA,Hattori_bulk,Sedarkian_bulk,Agasian_bulk1,Agasian_bulk2,Manu3,Manu4,Balbeer},
where shear viscosity~\cite{Tuchin,Li_shear,Asutosh,G_NJL_B,Nam,Tawfik,JD2,HRGB,HM3,Denicol,BAMPS}, electrical 
conductivity~\cite{Sedrakian_el,Kerbikov:2014ofa,Nam:2012sg,Hattori:2016lqx,Manu1,Manu2,Feng_cond,Fukushima_cond,Arpan1,Arpan2,NJLB_el,Lata,BC,Manu_SA},
bulk viscosity~\cite{Hattori_bulk,Sedarkian_bulk,Agasian_bulk1,Agasian_bulk2,Manu3}
for light quark sector
as well as field impact in heavy quark sector~\cite{Manu4,Balbeer} are studied.
%
%
Transport coefficients at finite magnetic field in the direction of 
gauge gravity duality is also studied in Refs.~\cite{ADS1,ADS2,mamo}.

In absence of magnetic field, any transport coefficient in different direction remain same i.e. an isotropic value is expected but in presence
of magnetic field, a multi-component structure of the transport coefficient can be built, for which an anisotropic transportation in the fluid is expected. 
Among the earlier references, which have not considered the rich multi-component structure of transport coefficients, will not be our matter of interest. They may be considered for rough estimations of magnetic field dependent transport coefficients, but they have missed the vital anisotropic part of transport properties in presence magnetic field. So for getting complete picture, we should focus only those particular references, who have considered the multi-component structure in well manner. Reviewing all those references, who considered the multi-component structure of transport coefficients, present article intended to provide a complete and enriched understanding on those multi-component transport coefficients, mainly shear viscosity along with electrical conductivity.

Now, if we analyze the existing microscopic calculations of shear viscosity in presence of magnetic 
field~\cite{Tuchin,Asutosh,G_NJL_B,JD2,HRGB,HM3,Denicol,BAMPS,ADS1,ADS2}, then we can classify them into two groups.
One group~\cite{Tuchin,Asutosh,G_NJL_B,JD2,HRGB,HM3} has considered a set of tensor structures, proposed by Ref.~\cite{Landau} and other group~\cite{Denicol,BAMPS,ADS1,ADS2} has considered another set of tensor structures, proposed by Ref.~\cite{XGH1,XGH2}. Therefore, we find two sets
of shear viscosities, which are also mutually inter-connected and can be linked with basic components like parallel, perpendicular and Hall components. Present article has attempted to provide the complete pictures of two set of shear viscosity components with their detail derivation in relaxation time approximation of kinetic theory approach. Although earlier Refs.~\cite{Tuchin,Asutosh,G_NJL_B,HRGB,Denicol} also addressed same final (simplified massless) expressions, but derivation of present work has its own uniqueness in its mathematical steps and approaches. After providing the complete picture two set of shear viscosity components with their general and then massless expressions, present work is also aimed to provide estimation first for three different distribution functions Maxwell-Boltzmann (MB), Fermi-Dirac (FD) and Bose-Einstein (BE), then for massless quark gluon plasma (QGP) and hadronic matter (HM) phases to visualize rough order of magnitude in two phases. At the end, lower bound
aspects of shear viscosity to entropy density ratio is discussed.

 
The article is organized as follows. Sec.~(\ref{eta_el_B}) covers a detail derivation of two sets of
shear viscosity components at finite magnetic field along with brief description of
electrical conductivity. Next in Sec.~\ref{sec_m0}, we have explored the massless case
results, where analytic expressions of different thermodynamical quantities and 
transport coefficients are obtained. Then in Sec.~(\ref{KSS_Bn0}),
we have explored the phenomenological band of viscosity to entropy density ratio for quark gluon plasma 
without and with magnetic field pictures. In Sec.~(\ref{sec:KSS}), the lower bound expectation of viscosity to entropy density ratio is discussed.
At the end, Sec.~(\ref{sum}) has summarized the investigations and few calculation gap
are addressed in Sec.~(\ref{App}).

We use the natural unit with $\hbar = k = c = 1$ and metric tensor $g^{\mu \nu} = \mathrm{diag} (+1,-1,-1,-1)$.
\section{Expressions of transport coefficients in presence of magnetic field}
\label{eta_el_B}
\subsection{Shear viscosity components using tensors of Ref.~\cite{Landau}}
\label{eta_L}
%

Before going to detail calculation of shear viscosity in presence of magnetic field, let us
take a quick revisit of its without field expression. There are two approaches -
Kubo framework~\cite{Jeon,Nicola,G_Kubo} and kinetic theory framework~\cite{chakrabortty,Gavin},
which are popularly practiced for microscopic calculations of transport coefficients. Here we will
follow the latter approach in relaxation time approximation (RTA).

Considering ideal ($T^{\mu\nu}_0$) and dissipative ($\Delta T^{\mu\nu}$) parts
of energy-momentum tensor as a macroscopic outcomes of two microscopic quantities - 
equilibrium distribution function $f_0$ and its deviation $\delta f$ (i.e. total distribution 
function $f = f_0 +\delta f$), one can express macroscopic
and microscopic definition of dissipative energy momentum tensor (shear viscosity part only) as
\bea
\Delta T^{\mu\nu} &=&\eta C^{\mu\nu} ~({\rm Macro})~,~
\nn\\
{\rm with}~ C^{\mu\nu} &=& \Big(\Delta^{\mu\alpha}\partial_\alpha u^\nu +  \Delta^{\nu\alpha}\partial_\alpha u^\mu -\frac{2}{3}
\Delta^{\mu\nu}\partial_\sigma u^\sigma \Big)~,
\nn\\
\Delta T^{\mu\nu}&=&g\int \frac{d^3\vk}{(2\pi)^3}\frac{k^\mu k^\nu}{\om}\delta f~({\rm Micro})~,
\label{TD_eta}
\eea
where $\eta$ is shear viscosity coefficient, $\Delta^{\mu\nu}=g^{\mu\nu}-u^\mu u^\nu$,
is the projection orthogonal to fluid velocity $u^\mu$, $g^{\mu \nu}$ is the metric tensor, $\om=\sqrt{\vk^2+m^2}$ is the energy of
fluid particle with mass $m$ and $g$ is its degeneracy factor. Assuming 
$\delta f=C\frac{k_\sigma}{\om} \frac{k_\rho}{\om} C^{\sigma\rho}$
in terms of an unknown constant $C$, Eq.~(\ref{TD_eta}) will provide the expression of $\eta$:
\be
\eta = \frac{2}{15}\int{\frac{d^3\vk}{(2\pi)^3}\: C \: \frac{\vk^4}{\om^3}}~,
\label{etaC_B0}
\ee
where the $1/15$ factor comes after using the identity (in 3-vector notation):
\be
\langle k^i k^j k^k k^l \rangle = \frac{1}{15}k^4 (\delta^{i j} \delta^{kl} + \delta^{il}\delta^{jk} + \delta^{ik}\delta^{jl})~.
\label{k_av}
\ee
With the help of the RTA in relativistic Boltzmann equation (RBE), the values of unknown constant can be obtained as
$C=-\frac{\beta\tau_c\om}{2}f_0(1-af_0)$~\cite{chakrabortty,Gavin}, where distribution function $f_0=1/\{e^{\beta\om}+a\}$
will be Fermi-Dirac (FD), Bose-Einstein (BE) and Maxwell-Boltzmann (MB) for $a=+1, -1, 0$.
Hence, Eq.~(\ref{etaC_B0}) will get the final form~\cite{chakrabortty,Gavin},
\be
\eta=\frac{g\beta}{15}\int \frac{d^3\vk}{(2\pi)^3}\frac{\vk^4}{\om^2}\tau_c f_0 (1-af_0)~.
\label{eta_B0}
\ee
Following the similar steps, adopted earlier for $B=0$ case~\cite{chakrabortty,Gavin},
we will now proceed for calculation of shear viscosity in presence of a magnetic field.
As viscosity is nothing to do with the time component, so, we can do the calculation by dropping
the time component also. For simplicity, in rest of the calculation we use 3-vector and
instead of Greek letters used for 4-vector in the above calculations we use {i,j,k,..} for 3-vector.
  
In presence of a magnetic field ($\vec B$) we can find 
5 independent traceless tensor components $C^n_{ij}$ ($n$ denotes different components which runs from $0$ to $4$)~\cite{Landau,Tuchin}
instead of a single tensor component $C_{ij}$. Hence, similar to Eq.~(\ref{TD_eta}),
the macroscopic to microscopic connection will be~\cite{Landau,Tuchin} (in 3-vector notation):
\bea
\Delta T_{i j} &=& \sum_{n=0}^{4} \eta_n C_{i j}^n ~({\rm Macro})
\nn\\
&=& g\int{\frac{d^3k}{(2\pi)^3}\: v_{i}v_{j} \: \omega \delta f}~({\rm Micro})
\label{TDth}
\eea
where $v_i\equiv \vk/\om~(=\vec v)$  is particle velocity, $\eta_n$ are 5 viscosity coefficients associated with 5
independent traceless tensors~\cite{Landau}:
\bea
C_{i j}^0 &=& (3b_i b_j - \delta_{i j}) (b_k b_l V_{k l} - (1/3)\vn \cdot \mathbf{V}),
\nn\\
C_{i j}^1 &=& 2V_{i j}+\delta_{i j}V_{k l}b_{k}b_{l}-2V_{i k}b_{k}b_{j}-2V_{j k}b_{k}b_{i} 
\nn\\
&&(b_i b_j - \delta_{i j})\vn \cdot \mathbf{V} + b_i b_j V_{k l}b_k b_l,
\nn\\
C_{i j}^2 &=& 2(V_{i k}b_j b_k + V_{j k}b_i b_k -2b_i b_j V_{k l}b_k b_l),  
\nn\\
C_{i j}^3 &=& V_{i k}b_{j k} + V_{j k}b_{i k}- V_{k l}b_{i k}b_j b_l - V_{k l}b_{j k}b_i b_l,  
\nn\\
C_{i j}^4 &=& 2(V_{k l}b_{i k}b_j b_l + V_{k l}b_{j k}b_i b_l)~.
\label{Cab}
\eea
Where, $V_{ij} = \frac{1}{2}\Big(\frac{\del V_i}{\del x_j}+\frac{\del V_j}{\del x_i}\Big)$
with fluid velocity $V_i\equiv \mathbf{V}$. $b_i\equiv {\vec B}/|B|$ is unit vector along the direction of magnetic field
${\vec B}$ and $b_{ij}=\varepsilon_{ijk}b_{k}$, $\varepsilon_{ijk}$ is total anti-symmetric Levi-Civita tensor.
The reason of changes from single possible traceless tensor $C^{ij}$ (at $B=0$) to multiple independent traceless tensors
$C_{ij}^n$ (at $B\neq 0$) is as follows. The $C^{ij}$ is built by $V^i$ and $\delta^{ij}$ only, but $C^n_{ij}$
has some additional building blocks $b_i$ and $b_{ij}$.

Similar to $B=0$ case, assuming $\delta f$ in terms of unknown constant $C_n$ (for $B=0$ case,
it was 1 unknown constant $C$, now for $B\neq 0$, it is 5 unknown constant $C_n$)~\cite{Landau,Tuchin}:
\bea
\delta f =\sum_{n=0}^{4} C_n C^n_{kl} v_{k} v_{l}~,
\label{delf}
\eea
Eq.~(\ref{TDth}) will give us expressions of $\eta_n$ in terms of $C_n$:
\bea
\eta_n = -\frac{2}{15}\int{\frac{d^3\vk}{(2\pi)^3}\: \omega \: C_n \: \vec v^4}~.
\label{etag}
\eea
With the help of RTA version of RBE, one can find the unknown constants $C_n$.
Refs.~\cite{Landau,Tuchin} have found these unknown constants in strong field
limit. Present work has found its general expressions, valid for all ranges of
magnetic field. Refs.~\cite{Asutosh,HRGB} have also found similar kind of general
expressions, whose final expressions are exactly same as obtained in present draft,
although reader can identify the differences in mathematical steps. Ref.~\cite{Asutosh} 
has explored the root of the traceless independent tensors $C^n_{ij}$ from the 
simple possible form of independent tensors with building blocks $\delta_{ij}$, $b_i$
and $b_{ij}$. While Ref.~\cite{HRGB} has gone through the similar mathematical structure via 
projection operator technique. In this context, present calculations show its quick and 
easy path for those reader, who are acquainted with the 5 possible independent traceless
tensors $C_{ij}^n$. By repeating the similar steps of $B=0$ case, prescribed in Refs.~\cite{Gavin,chakrabortty} one can easily find
the respective coefficients $\eta_n$, linked with $C_{ij}^n$. Present article has basically explored it.

Now, let us proceed to find $C_n$ from RTA based RBE in presence of magnetic field, whose mathematical form looks like:
\bea
&&\frac{1}{\om}p_{i}\frac{\del f}{\del x_{i}} 
+ \frac{1}{\om}qF_{ij}p_{i}\frac{\del f}{\del p_{j}} = -\frac{\delta f}{\tau_c}
\nn\\
&&-\frac{\omega}{T} v_{i}v_{j}V_{ij}f_0(1-a f_0) -\frac{qB}{\omega}b_{ij}v_{j}\frac{\del}{\del v_{i}}(\delta f)
=  -\frac{\delta f}{\tau_c}
\label{RBE}
\eea
where electromagnetic field-strength tensor $F_{ij} = -B b_{ij}$ is applied on $q$ charge. Through this second term of left hand side (LHS), magnetic field enters into the calculation. 
LHS of RBE is considered $f\approx f_0$, where $\frac{\partial f_0}{\partial x_{i}}$ will give $1^{\rm st}$
term of lhs but $2^{\rm nd}$ term will be vanished. For getting non-zero values of $2^{\rm nd}$ term,
which will basically consist of B-dependent term, we have to consider $\delta f$ correction term.
Hence $2^{\rm nd}$ term of lhs carry $\frac{\del (\delta f)}{\del v_j}$.
Similar to relaxation time $\tau_c$, we will get another time scale $\tau_B=\om/(qB)$ in Eq.~(\ref{RBE}),
which is similar to inverse of cyclotron frequency.
Former is mainly controlled by randomness of the medium at particular temperature $T$, but it may also 
have dependence with magnetic field. 
Although in present work, we will consider it as free parameter. The $\tau_B$ is completely originated 
from external magnetic field and inversely depends on it.

Now by using $\delta f$ from Eq.~(\ref{delf}) in Eq.~(\ref{RBE}), we get 
\begin{eqnarray}
\nn \frac{\omega}{T} v_{i}v_{j}V_{ij}f_0(1-a f_0)=-\frac{1}{\tau_B}b_{ij}v_{j}\frac{\partial}{\partial v_{i}}(\sum_{n=0}^{4} C_n C^n_{kl} v_{k} v_{l})&&\\
\nn
 + \frac{1}{\tau_c}(\sum_{n=0}^{4} C_n C^n_{kl} v_{k} v_{l})&&\\
\nn \Rightarrow \frac{\omega}{T} v_{i}v_{j}V_{ij}f_0(1-a f_0) = -\frac{1}{\tau_B}b_{ij}v_{j}\:2\times(\sum_{n=0}^{4} C_n C^n_{ik} v_{k})&&\\
 + \frac{1}{\tau_c}(\sum_{n=0}^{4} C_n C^n_{kl} v_{k} v_{l})~~~&&
 \label{RBE1}
\end{eqnarray}

Now, comparing same tensor structure on both side (whose details calculations are given in Appendix, labeled as 
Sec~.(\ref{App1_L})) we get,
\bea
\nn
C_1&=&-\frac{\omega}{2T}\;
\frac{\tau_c}{4\left\{\frac{1}{4}+(\tau_c/\tau_B)^2\right\}} ~ f_0(1-a f_0)\\
\nn
C_2&=&-\frac{\omega}{2T}\; \frac{\tau_c}{1+(\frac{\tau_c}{\tau_B})^2} ~ f_0(1-a f_0)\\
\nn
C_3&=&-\frac{\omega}{2T}\; \frac{\tau_c(\frac{\tau_c}{\tau_B})}{2\left\{\frac{1}{4}+(\tau_c/\tau_B)^2\right\}}
~ f_0(1-a f_0)\\
C_4&=&-\frac{\omega}{2T}\; \frac{\tau_c(\frac{\tau_c}{\tau_B})}{1+(\frac{\tau_c}{\tau_B})^2} ~ f_0(1-a f_0)
\label{C's}
\eea
 
Now, using expressions of $C$s from eq.~(\ref{C's}) in eq.~(\ref{etag}) we get,
\bea
\nn
\eta_1&=&\frac{g\beta}{15}\int \frac{d^3\vk}{(2\pi)^3}\left(\frac{\vk^2}{\om}\right)^2\tau_c
\frac{1}{4\left\{\frac{1}{4}+(\tau_c/\tau_B)^2\right\}}
\\ \nn
&&\left\{f_0(1-a f_0)\right\}
\label{eta1_B}
\\
\nn
\eta_2&=&\frac{g\beta}{15}\int \frac{d^3\vk}{(2\pi)^3} \left(\frac{\vk^2}{\om}\right)^2\tau_c
\frac{1}{1+(\tau_c/\tau_B)^2}
\\ \nn
&&\left\{f_0(1-a f_0)\right\}
\label{eta2_B}
\\
\nn
\eta_3&=&\frac{g\beta}{15}\int \frac{d^3\vk}{(2\pi)^3}\left(\frac{\vk^2}{\om}\right)^2 \tau_c
\frac{\tau_c/\tau_B}{2\left\{\frac{1}{4}+(\tau_c/\tau_B)^2\right\}}
\\ \nn
&&\left\{f_0(1-a f_0)\right\}
\label{eta3_B}
\\
\nn
\eta_4&=&\frac{g\beta}{15}\int \frac{d^3\vk}{(2\pi)^3}\left(\frac{\vk^2}{\om}\right)^2 \tau_c
\frac{\tau_c/\tau_B}{1+(\tau_c/\tau_B)^2}
\\
&&\left\{f_0(1-a f_0)\right\}.
\label{eta4_B}
\eea
Now, one can find $C^0_{ij}=0$ for zero bulk viscosity case where
$\mathbf{\nabla} \cdot \mathbf{V}=V_{ii}=0$, and
$V_{ij}b_{i}b_{j}=0$. However $\eta_0$ might have a non-zero value,
which may not obtained via earlier methodology, so we might have to find some alternative way. From eq.~(\ref{Cab}) we can find that $C^0_{ij}$ is parallel to
magnetic field because $C^0_{ij}b_i b_j \neq 0$. As in the field direction Lorentz force
has no effect, so, $\eta_0$ should be same as in zero magnetic field ($B=0$), 
which is given in eq.~(\ref{eta_B0})
\be
\eta_0 = \eta =\frac{g\beta}{15}\int \frac{d^3\vk}{(2\pi)^3}\frac{\vk^4}{\om^2}\tau_c f_0 (1-af_0)~.
\ee
Our conclusion can be checked by alternative calculations, done in Refs.~\cite{Asutosh,HRGB}.

{\bf Limiting Cases:}
These four 
different viscosity components can be treated as more general expressions, 
which can able to generate the expressions of strong field limit, addressed in Refs.~\cite{Landau,Tuchin} as a special case of them.
Two possible limiting cases are 
discussed below.
\begin{figure}
\resizebox{0.45\textwidth}{!}{
\includegraphics{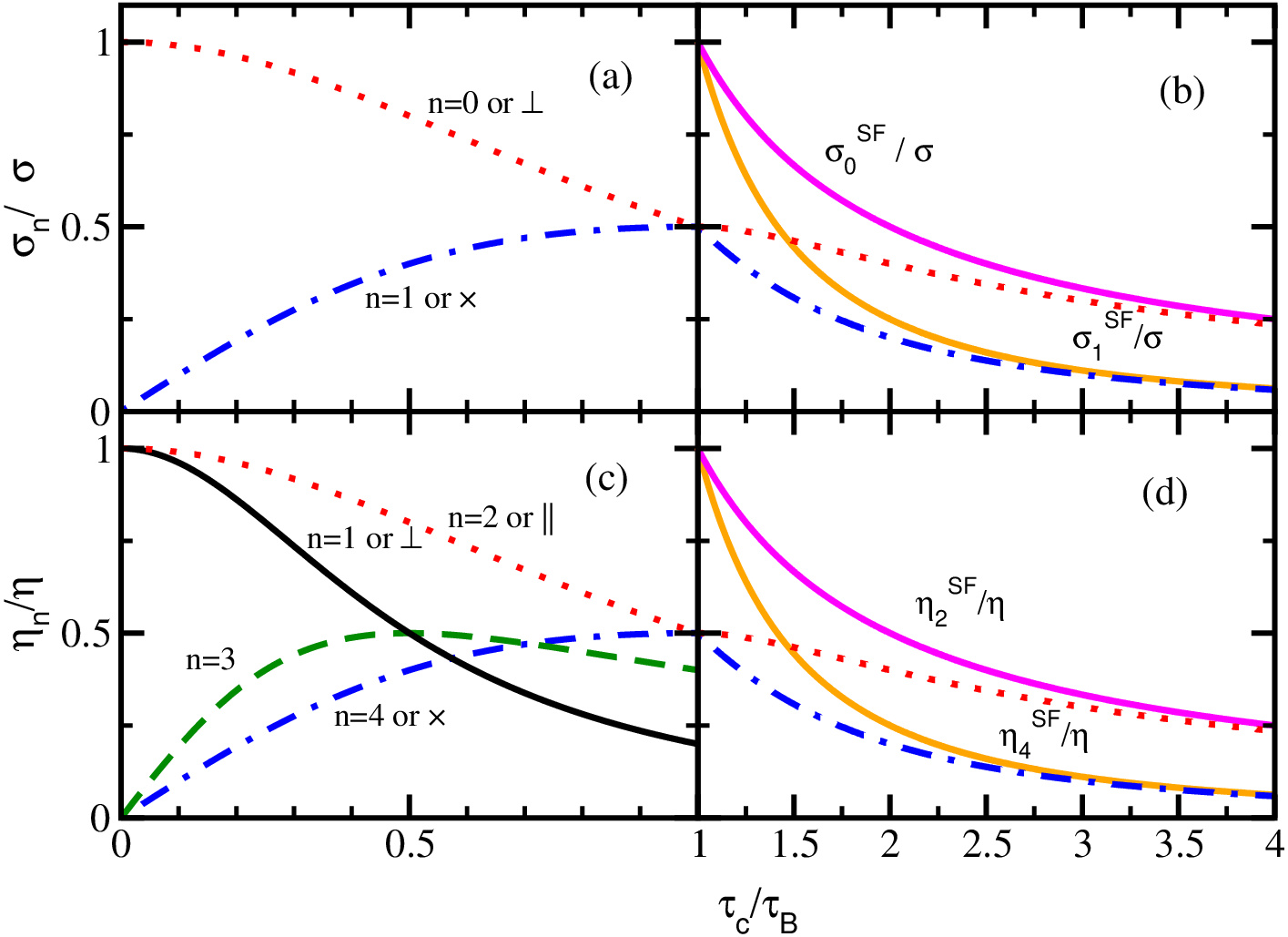}
}
\caption{Ratio of conductivity (a,b) and viscosity (c,d) with and without magnetic field 
along $\frac{\tau_c}{\tau_B}$ axis, which is classified into two: $\frac{\tau_c}{\tau_B}<1$ (a,c)
and $\frac{\tau_c}{\tau_B}>1$ (b,d).}
\label{fig:sh_tcB}
\end{figure}

For $B\rightarrow 0$, $\tau_B\rightarrow \infty$, we
get
\bea
\eta_2&=&\eta_1
\nn\\
&=&\frac{g\beta}{15}\int \frac{d^3\vk}{(2\pi)^3}\tau_c
\left(\frac{\vk^2}{\om}\right)^2 [f_0\{1-a f_0\}]
\nn\\
&=&\eta
\label{eta_2B0}
\eea
and
\be
\eta_4= \eta_3= 0~.
\ee
One can identify $\eta_{1,2}$ as normal shear viscosity as they merge to $\eta$
at $B\rightarrow 0$.
Seeing the vanishing values of $\eta_{3,4}$ in absence of magnetic field, one can realize them
as Hall-type shear viscosity, as they completely originate 
from $qB$ and odd function of $qB$. 
So, let us relate them with other notations like $\eta_{\parallel,\perp}$, used in earlier Refs.~\cite{ADS1,ADS2}, based on holographic dual theory.
If the magnetic field is applied along z direction, then 
$\eta_{xyxy}$ and $\eta_{xzxz}=\eta_{yzyz}$ can be called as perpendicular and
parallel components. We have identified the relations:
\bea
\eta_{xyxy}=\eta_{\perp}&=&\eta_1
\nn\\
\eta_{xzxz}=\eta_{yzyz}=\eta_{\parallel}&=&\eta_{2}~.
\label{pp_L}
\eea
Sometimes the Hall viscosity components $\eta_{3,4}$ is
denoted as $\eta_{\times}$~\cite{HRGB}.

If we take strong field (SF) limit $B\rightarrow \infty$, for which $\tau_B\rightarrow 0$,
then we will get
\bea
\eta_2&=&4\eta_1
\nn\\
&=&\frac{g\beta}{15}\int \frac{d^3\vk}{(2\pi)^3}\frac{\tau_B^2}{\tau_c}
\left(\frac{\vk^2}{\om}\right)^2[f_0\{1-a f_0\}]
\nn\\
&=&\eta^{SF}_{2} ({\rm say})
\label{eta_2tB0}
\eea
and
\bea
\eta_4&=&2\eta_3
\nn\\
&=&\frac{g\beta}{15}\int \frac{d^3\vk}{(2\pi)^3}\tau_B
\left(\frac{\vk^2}{\om}\right)^2[f_0\{1-a f_0\}]
\nn\\
&=& \eta^{SF}_{4} ({\rm say})~.
\label{eta_4tB0}
\eea
Exactly same expressions have been obtained 
in Refs.~\cite{Tuchin,Landau} in terms of MB distribution, where the calculations
are done in strong field assumption.

\subsection{Shear viscosity components using tensors of Ref.~\cite{XGH1,XGH2}}
\label{eta_R}
%
 %
In this section, we will find other set of shear viscosity
components $\eta_n$ if we start with other set of independent
traceless tensors ($C^n_{ij}$) instead previous set, given in Eq.~(\ref{Cab}).
This other set of tensors is proposed in Refs.~\cite{XGH1,XGH2}
and their structures are given below:
\bea
\nn 
C^0_{ij} &=& 2 \left(V_{ij} + (1/3) \delta_{ij} \vn \cdot \vV \right) \\
\nn
C^1_{ij} &=&\frac{3}{4} \left\{\left(3b_i b_j - \delta_{i j}\right) \left(b_k b_l V_{k l} - (1/3) \vn \cdot \mathbf{V}\right)\right\}\\
\nn	
C^2_{ij} &=& 2 \left(V_{ik}b_j b_k + V_{jk}b_i b_k - 2V_{ij} b_i b_j b_k b_l \right) \\
\nn C^3_{ij} &=& 2\left( V_{ik} b_{jk} + V_{jk} b_{ik} - V_{kl} b_{jl} 
b_{i} b_{k} - V_{kl} b_{il} b_{j} b_{k} \right)\\
\nn	C^4_{ij} &=& 2\left( b_{ik} b_j b_l + b_{jk} b_i b_l
\right) V_{kl} \\
\label{Cmn_R}
\eea
All notations have same meaning as in the previous section.

Now, using $C_{ij}$s of Eq.~(\ref{Cmn_R}) in Eq.~(\ref{RBE1}),
one can find all $C_n$ as done for earlier subsection. The details of $C_n$ calculations are addressed in Appendix - Sec.~(\ref{App1_R}).
Using those $C_n$'s, we get the final expressions of $\eta_n$ as,
\bea
\eta_0&=&\frac{g\beta}{15}\int \frac{d^3\vk}{(2\pi)^3}\left(\frac{\vk^2}{\om}\right)^2\tau_c
\frac{1}{\left\{1+4(\tau_c/\tau_B)^2\right\}}
\nn\\
&& \left\{f_0(1-a f_0) \right\},
\label{eta0_BR}
\nn\\
\eta_2&=&\frac{3g\beta}{15}\int \frac{d^3\vk}{(2\pi)^3}\left(\frac{\vk^2}{\om}\right)^2\tau_c
\frac{(\tau_c/\tau_B)^2}{\{1+4(\tau_c/\tau_B)^2\}\{1+(\tau_c/\tau_B)^2\}} 
\nn\\
&&\left\{f_0(1-a f_0) \right\},
\label{eta2_BR}
\nn\\
\eta_3&=&\frac{g\beta}{15}\int \frac{d^3\vk}{(2\pi)^3}\left(\frac{\vk^2}{\om}\right)^2\tau_c
\frac{(\tau_c/\tau_B)}{\{1+4(\tau_c/\tau_B)^2\}}
\nn\\ 
&&  \left\{f_0(1-a f_0) \right\},
\label{eta3_BR}
\nn\\
\eta_4&=&\frac{g\beta}{15}\int \frac{d^3\vk}{(2\pi)^3}\left(\frac{\vk^2}{\om}\right)^2\tau_c
\frac{(\tau_c/\tau_B)}{\{1+(\tau_c/\tau_B)^2\}}
\nn\\
&& \left\{f_0(1-a f_0) \right\}~.
\label{eta4_BR}
\eea
Now, only $\eta_1$ can not be obtained from RBE, so
we will find some alternative way to obtain its expression.
We see that Among the earlier set of $C_{ij}^n$ in Eq.~(\ref{Cab}),
only $C^0_{ij}$ is parallel to magnetic field ($C^0_{ij} b_i b_j \neq 0$). Similarly, for present set of $C_{ij}^n$ in Eq.~(\ref{Cmn_R}), $C^0_{ij}$ and $C^1_{ij}$ are components of shear-stress tensor parallel
to magnetic field. However, we don't find the expressions
of $\eta_0$ and $\eta_1$ are equal to $\eta$ but if we
take linear combination of $\eta_0$ and $\frac{3}{4}\eta_1$ and demand
\be
\eta_0 + \frac{3}{4}\eta_1 = \eta~,
\ee
then we can get 
\bea
 \eta_1 &=& \frac{16}{3}\frac{g\beta}{15}\int \frac{d^3\vk}{(2\pi)^3}\frac{\vk^4}{\om^2} \tau_c \frac{(\tau_c/\tau_B)^2}{\{1+4(\tau_c/\tau_B)^2\}} \\
&& f_0 (1-af_0)~,
\eea 
which is exactly same (massless expressions), as obtained
by Ref.~\cite{Denicol}. In latter section, we will find
that our massless expressions of $\eta_n$ (obtained from
RTA based kinetic theory) are exactly same as obtained by
Ref.~\cite{Denicol} (in RTA based methods of moment technique).

When we compare present set of coefficients with earlier coefficients, we can identify two categories - one category
is already existed at $B=0$ picture, while other category
does not. It means that the coefficients of latter category
are completely appeared due to magnetic field. Hence, they can be called as magnetically induced coefficients. To distinguish
between two sets of coefficients, let us modify the notations of
earlier coefficient of Sec.~(\ref{eta_L}) as 
${\tilde\eta}_{0,1,2,3,4}$ from here.

One can easily realize that similar to ${\tilde\eta}_{3,4}$, the
$\eta_{1,2,3,4}$ are magnetically induced coefficients as
all are disappeared at $B\rightarrow 0$ or $\tau_B\rightarrow \infty$. Among them ${\tilde\eta}_{3,4}$ and 
${\eta}_{3,4}$ are Hall coefficients, which are odd function of $\tau_B$ or $eB$. They are connected as
\bea
\eta_3 &=& \frac{{\tilde \eta}_3}{2}
\nn\\
\eta_4 &=& {\tilde\eta}_4~.
\eea
One can now find similarities between $\eta_0$
and ${\tilde\eta}_{1,2}$ as these are not vanished
at $B=0$. They are connected as
\bea
\eta_0 &=& {\tilde \eta}_1
\nn\\
\eta_0 +\eta_2 &=& {\tilde\eta}_2~.
\eea
Remembering the $\eta_{\parallel,\perp}$
components in Eq.~(\ref{pp_L}), we can identify
the relations:
\bea
\eta_{\perp}=\eta_0 &=& {\tilde \eta}_1
\nn\\
\eta_{\parallel}=\eta_0 +\eta_2 &=& {\tilde\eta}_2~.
\eea
Remaining coefficients are connected as
\bea
\eta_2 &=& {\tilde \eta}_2 - {\tilde \eta}_1
\nn\\
\eta_0 +\frac{3}{4}\eta_1 &=& {\tilde\eta}_0=\eta~.
\eea
So instead of being confused with too many coefficients,
we should be focus on the gross physics, which is as follows. In absence of magnetic field, shear viscosity
in different direction remain same and this isotropic
value is defined as $\eta$ here. Now, in presence
of magnetic field this isotropic property breaks and we
get different values of $\eta_{\parallel}$ and $\eta_{\perp}$, which will be merged to $\eta$ at $B\rightarrow 0$ i.e. 
\be
\lim_{B\rightarrow 0}\eta_{\parallel}\rightarrow \eta \leftarrow \lim_{B\rightarrow 0}\eta_{\perp}~. 
\ee
In terms of two sets of coefficients, we can write as
\bea
\lim_{B\rightarrow 0}{\tilde\eta}_2\rightarrow
&{\tilde\eta}_0& \leftarrow \lim_{B\rightarrow 0}{\tilde\eta}_1,
\nn\\
\lim_{B\rightarrow 0}(\eta_0 + \eta_2)\rightarrow &\Big(\eta_0 +\frac{3}{4}\eta_1\Big)& \leftarrow\lim_{B\rightarrow 0}\eta_0~.
\eea
There also be Hall viscosity $\eta_{\times}$, which is
absent at $B\rightarrow 0$ and that can be mathematically represented as
\bea
\lim_{B\rightarrow 0}\eta_{\times}&\rightarrow & 0
\nn\\
\lim_{B\rightarrow 0}{\tilde\eta}_{3,4}&\rightarrow &0
\nn\\
\lim_{B\rightarrow 0}\eta_{3,4}&\rightarrow & 0~.
\eea


\subsection{Electrical conductivity components}
\label{sec:el}
Next, we come to the expression of electrical conductivity in presence of
the magnetic field. Here, again we can first recall without magnetic field
expression of the electrical conductivity,
\be
\sigma = e^2 g {\beta} \, \frac{1}{3} \int \frac{d^3\vk}{(2\pi)^3}\, 
\frac{{\vk}^2}{\om^2} \tau_c f_0(1-f_0),
\label{cond_B0}
\ee
where, $e$ is elementary charge and all other symbols have same meaning as 
in the above subsections. Then, instead of repeating its formalism
for $\vec B \ne 0$ case, given in Refs.\cite{Sedrakian_el},
let us come directly to final expressions
\be
\sigma_n = e^2 g {\beta} \, \frac{1}{3} \int \frac{d^3\vk}{(2\pi)^3}\, 
\frac{{\vk}^2}{\om^2} 
\frac{\tau_c(\tau_c/\tau_B)^n}{1+(\tau_c/\tau_B)^2} f_0(1-f_0),
\label{cond_n}
\ee

where, $n$ can take any value from $0$ to $2$, and $\sigma_0$ is normal conductivity along $xx$ or $yy$ direction, $\sigma_1$ is Hall conductivity
along $xy$ or $yx$ direction and $\sigma_{zz}=\sigma_0+\sigma_2=\sigma$ is longitudinal conductivity
if $B$ is applied along $z$-direction. Alternative 
notations $\sigma_{\perp}=\sigma_0$, 
$\sigma_{\parallel}=\sigma_{0} +\sigma_2$ and $\sigma_{\times}=\sigma_1$ are also used in earlier
references~\cite{HRGB,Arpan1,Arpan2}.

Here also, magnetic field create anisotropic conduction
($\sigma_{\perp}\neq\sigma_{\parallel}$) and a magnetically induced Hall conductivity $\sigma_{\times}$.
At $B\rightarrow 0$, restoring the isotropic property and vanishing
Hall conductivity can be mathematical expressed as
\bea
\lim_{B\rightarrow 0}\sigma_{\perp}&\rightarrow &\sigma_{\parallel}=\sigma
\nn\\
\lim_{B\rightarrow 0}\sigma_{\times}&\rightarrow & 0~.
\eea
In strong field limit $B\rightarrow\infty$,
we will get:
\bea
\sigma_0&=&\frac{g\beta e^2}{3}\int \frac{d^3\vk}{(2\pi)^3}\frac{\tau_B^2}{\tau_c}
\left(\frac{\vk^2}{\om}\right)^2[f_0\{1-a f_0\}]
\nn\\
&=&\sigma^{SF}_0 ({\rm say})
\label{con_0tB0}
\eea
and
\bea
\sigma_1&=&\frac{g\beta e^2}{3}\int \frac{d^3\vk}{(2\pi)^3}\tau_B
\left(\frac{\vk^2}{\om}\right)^2[f_0\{1-a f_0\}]
\nn\\
&=& \sigma^{SF}_1 ({\rm say})
\label{con_1tB0}
\eea
Analytic outcomes of $\eta_{\parallel,\perp,\times}$ and $\sigma_{\parallel,\perp,\times}$ for two opposite limits
are drawn in Fig.~\ref{fig:sh_tcB}(a-d) for getting better 
visualization. $B\rightarrow 0$ and $B\rightarrow\infty$ can be 
alternatively realized by $\tau_c/\tau_B\rightarrow 0$ and 
$\tau_c/\tau_B\rightarrow\infty$ or $\tau_c/\tau_B<<1$ and 
$\tau_c/\tau_B>> 1$ in numerical point of view.
In Fig.~\ref{fig:sh_tcB}(a) and (b), $\sigma_{\parallel,\perp,\times}$ and $\eta_{\parallel,\perp,\times}$
are plotted against $\tau_c/\tau_B$ from 0 to 1, where we can find the 
merging $\sigma_{\perp}$ to $\sigma_{\parallel}=\sigma$ and $\eta_{\parallel,\perp}$ to $\eta$ at $\tau_c/\tau_B=0$.
It means that we can get back our isotropic values of conductivity ($\sigma$) and shear viscosity ($\eta$) in absence
of magnetic field just by putting $B\rightarrow 0$ or $\tau_c/\tau_B\rightarrow 0$.
On the other hand, Hall conductivity $\sigma_{\times}$ and shear viscosity components $\eta_{\times}$
is disappeared at $B\rightarrow 0$ or $\tau_c/\tau_B\rightarrow 0$ as it is completely
magnetic field induced phenomena. Next, in Fig.~\ref{fig:sh_tcB}(c) and (d), $\sigma_{\perp,\times}$ 
$\eta_{\perp,\times}$ are extended for $\tau_c/\tau_B> 1$ with same line-style curves. By using approximation
$\tau_c/\tau_B>>1$, we have already got strong field limit expressions of $\sigma_{\perp,\times}$ 
$\eta_{\perp,\times}$, which are renamed as $\sigma^{SF}_{\perp,\times}$ in Eqs.~(\ref{con_0tB0}), 
(\ref{con_1tB0}) and $\eta^{SF}_{\perp,\times}$ in Eqs.~(\ref{eta_2tB0}), (\ref{eta_4tB0}).
Plotting them against $\tau_c/\tau_B$-axis, we notice that $\sigma_{\perp,\times}$,
$\eta_{\perp,\times}$ are merging to $\sigma^{SF}_{\perp,\times}$, and $\eta^{SF}_{\perp,\times}$ around and beyond 
$\tau_c/\tau_B=4$. It means that we can safely use strong field approximated
expressions for $\tau_c/\tau_B\geq 4$ but they can not be used for 
$\tau_c/\tau_B<4$. 
\section{For massless Bosonic and Fermionic matter}
\label{sec_m0}
\subsection{Thermodynamics for $B=0$}
\label{Th_B0}
Here, we will address the analytic forms of thermodynamical quantities
like energy density $\epsilon$, pressure $P$, entropy density $s$ for different
systems of massless particles following Maxwell-Boltzmann (MB), Bose-Einstein (BE) and 
Fermi-Dirac (FD) distribution.
To estimate $\eta/s$ in Sec.~(\ref{KSS_Bn0}), $s$ is our main required quantity here.

In terms of distribution function $f_0$, the energy density and 
pressure of any medium can be expressed as
\bea
\epsilon &=& g \int_{0}^{\infty}\frac{d^3\bp}{(2\pi)^3} \: \om \: f_0~,
\nn\\
P &=& g \int_{0}^{\infty}\frac{d^3\bp}{(2\pi)^3} \: \frac{\bp^2}{3\om} \: f_0~,
\label{en_CM}
\eea
which are connected as $P = \frac{1}{3}\epsilon$ for massless case.
So, entropy density of the system is 
\bea
\nn s &=& \frac{\epsilon + P}{T} = \frac{4\epsilon}{3T}\\
&=&\frac{4g\beta}{3}  \int_{0}^{\infty}\frac{d^3\bp}{(2\pi)^3} \: \om \: f_0
\label{s_mb}
\eea
%

Solving Eq.~({\ref{s_mb}}) with $\vp = \om$ as for massless case, we get (See Sec.~(\ref{Ap_Th_B0}) in Appendix)
\bea
s &=& \frac{4g}{\pi^2} \: T^3~~{\rm for~MB}
\nn\\
&=& \frac{4g}{\pi^2}\zeta(4) \: T^3=\frac{4g\pi^2}{90} \: T^3 ~~{\rm for~BE}
\nn\\
&=& \Big(\frac{7}{8}\Big)\frac{4g}{\pi^2}\zeta(4) \: T^3=\frac{7g\pi^2}{180} \: T^3~~{\rm for~FD}~,
\label{s_anl}
\eea
where 
\bea
\zeta(4)&=&\frac{1}{\Gamma(4)}\int_0^\infty x^3/(e^x-1)
\nn\\
&=&\sum_{n=1}^\infty \frac{1}{n^4}=\frac{\pi^4}{90}~.
\eea

\subsection{Shear viscosity and electrical conductivity for $B=0$}
\label{Shel_B0}
  \begin{figure}
\resizebox{0.45\textwidth}{!}{
 	\includegraphics{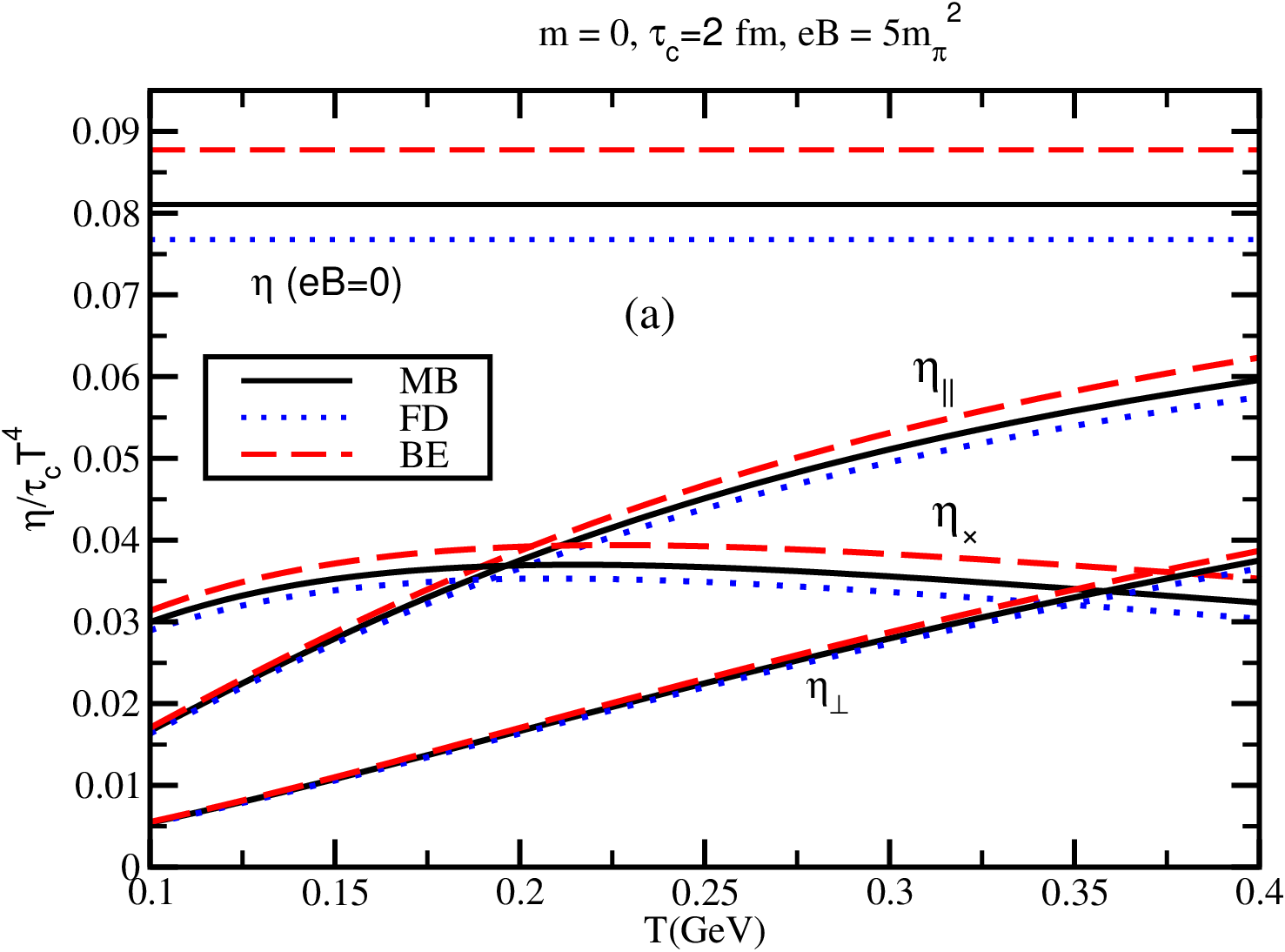}
 }
 \resizebox{0.45\textwidth}{!}{
 	\includegraphics{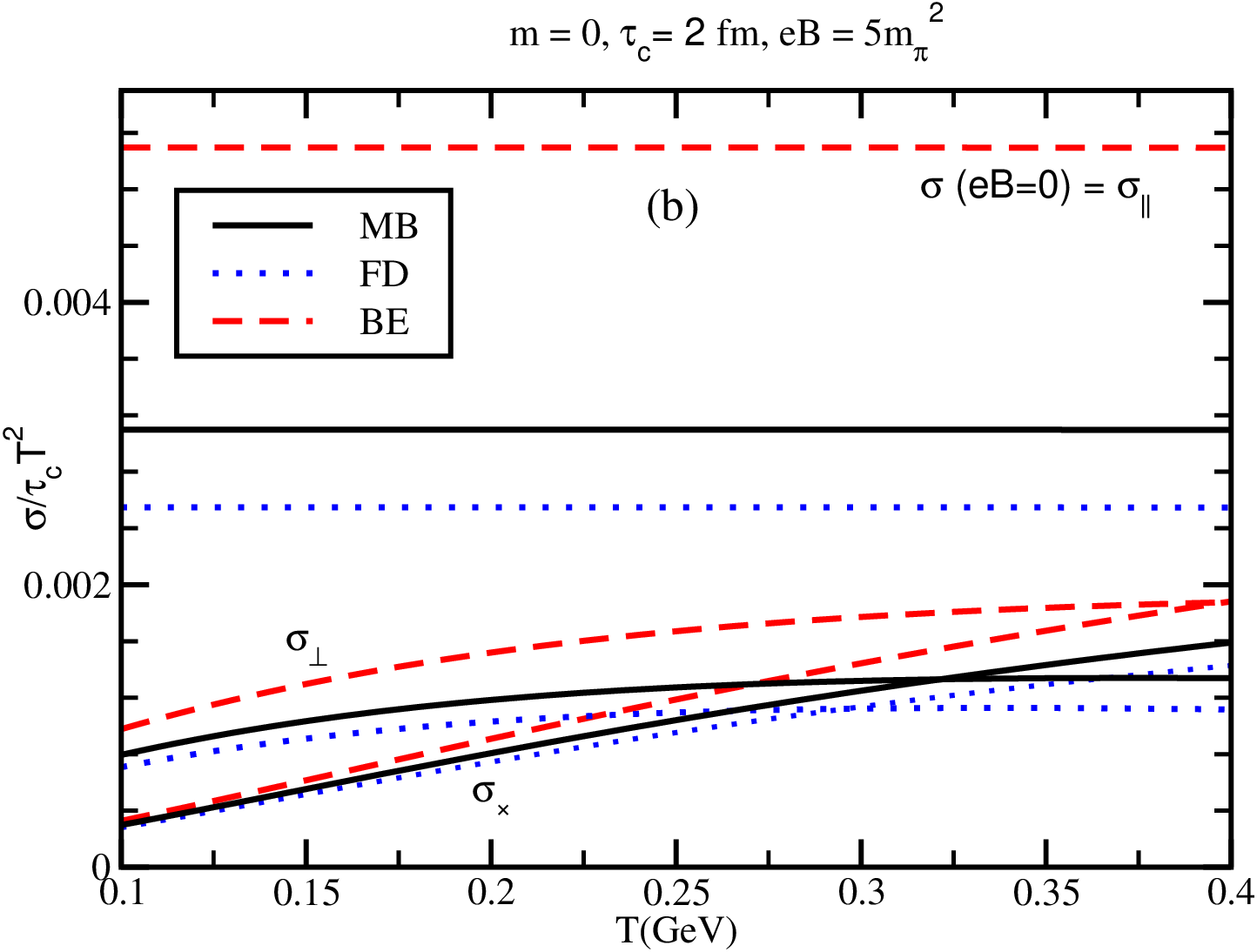}
 }
 \resizebox{0.45\textwidth}{!}{
 	\includegraphics{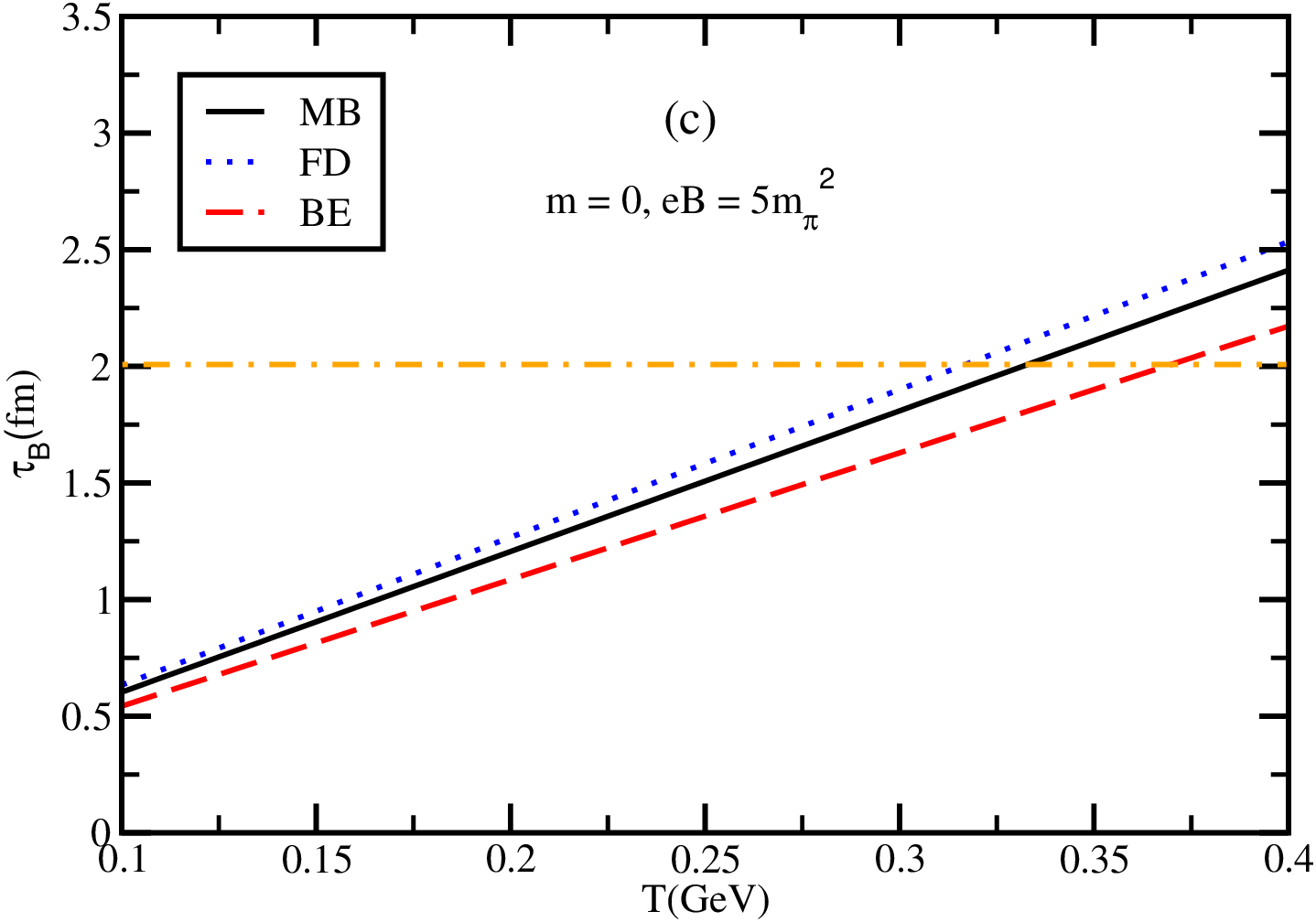}
 }	
 	\caption{$\eta/(\tau_cT^4)$, $\eta_{\parallel}/(\tau_cT^4)$, $\eta_{\perp}/(\tau_cT^4)$, $\eta_{\times}/(\tau_cT^4)$ vs $T$ (upper panel) $\sigma/(\tau_cT^2)$, $\sigma_{\perp}/(\tau_cT^2)$, $\sigma_{\times}/(\tau_cT^2)$ 
 	vs $T$ (middle panel) for MB, FD and BE distributions.
 	Lower panel: magnetic time scale for MB, FD and BE distributions}
 	\label{fig:estB_T}
  \end{figure}
 \begin{figure}
 \resizebox{0.45\textwidth}{!}{
 	\includegraphics{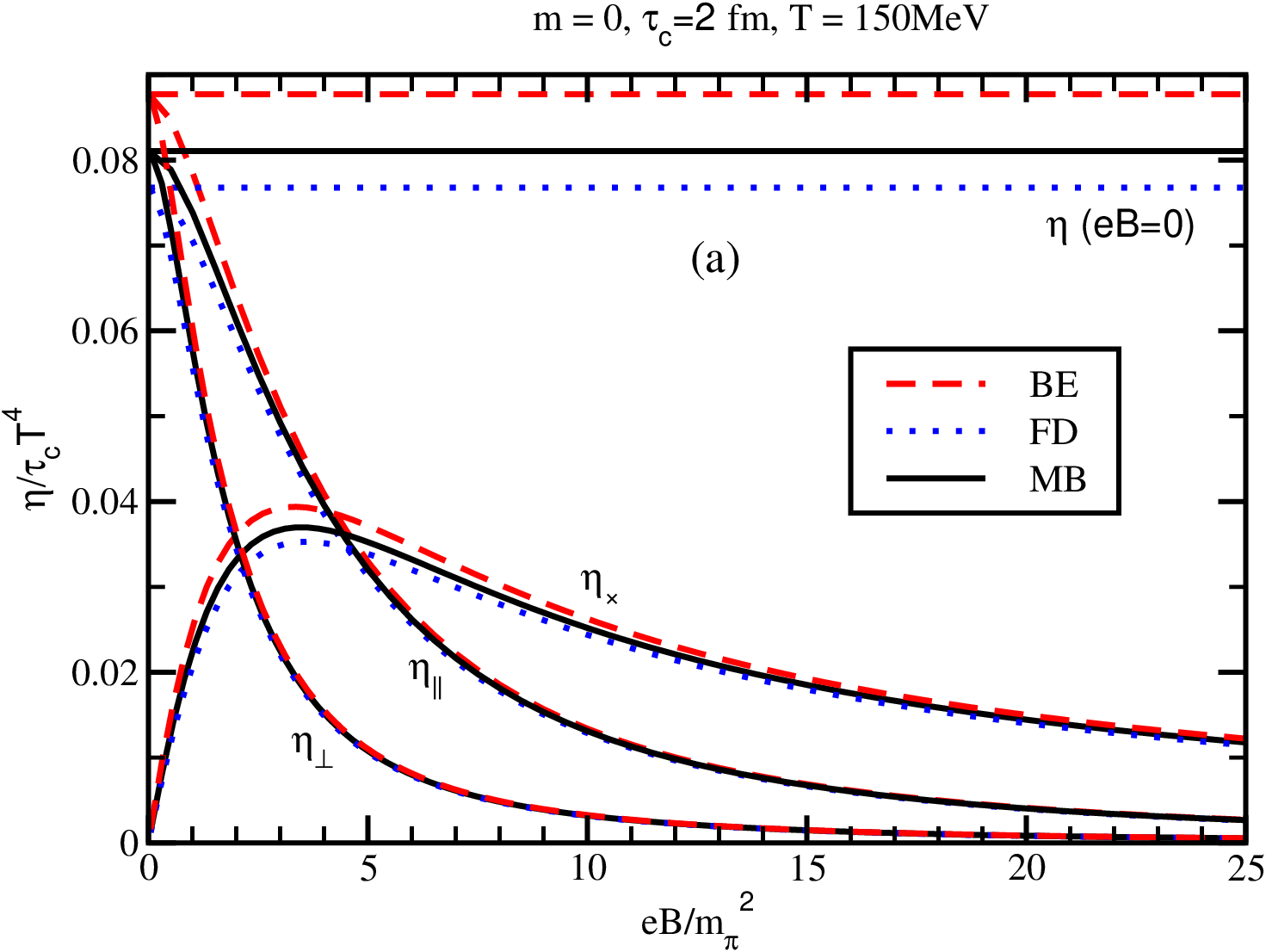}
 }
 \resizebox{0.45\textwidth}{!}{
 	\includegraphics{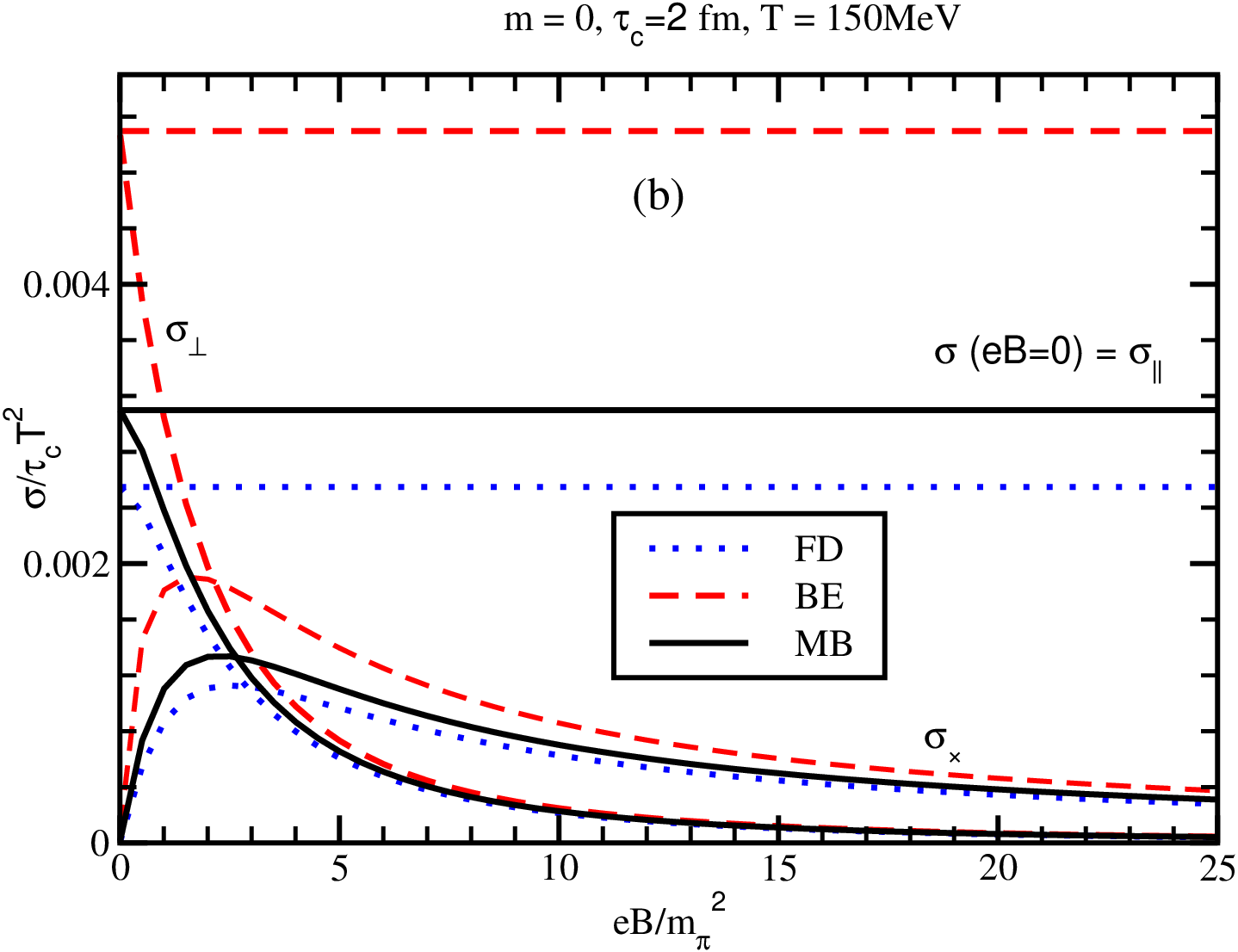}
 }
 \resizebox{0.45\textwidth}{!}{
 	\includegraphics{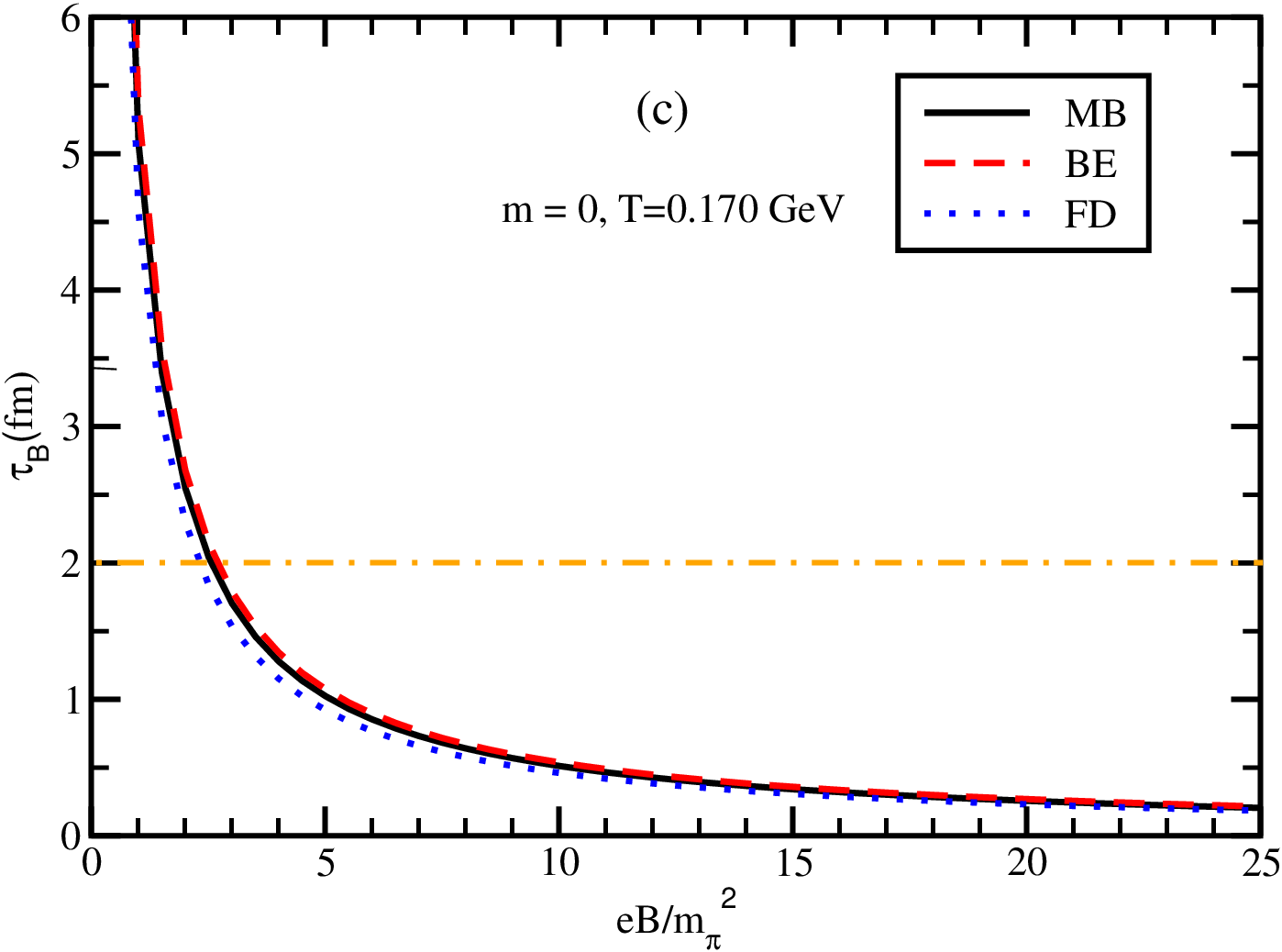}
 }
 	\caption{Same as Fig.~(\ref{fig:estB_T}) along $eB/m_\pi^2$-axis}
 	\label{fig:estB_B}
  \end{figure}
 Now, let us come to shear viscosity and electrical conductivity expressions
 for massless and $B=0$ case.
By solving Eq.~(\ref{eta_B0}) for massless case, one can get (See Sec.~(\ref{Ap_Shel_B0}) in Appendix):
\bea
\eta &=& \frac{4g\:\tau_c}{5\pi^2} \: T^4
~~{\rm for~MB}
\nn\\
&=& \frac{4g\:\tau_c}{5\pi^2}\zeta(4) \: T^4=\frac{4g\pi^2\:\tau_c}{450} \: T^4 ~~{\rm for~BE}
\nn\\
&=& \Big(\frac{7}{8}\Big)\frac{4g\:\tau_c}{5\pi^2}\zeta(4) \: T^4=\frac{7g\pi^2\:\tau_c}{900} \: T^4~~{\rm for~FD}
\label{eta_anl}
\eea
and solving Eq.~(\ref{cond_B0}) for massless case, we get:
\bea
\sigma &=& \frac{ge^2\:\tau_c}{3\pi^2} \: T^2
~~{\rm for~MB}
\nn\\
&=& \frac{ge^2\:\tau_c}{6\pi^2}\zeta(2) \: T^2=\frac{ge^2\:\tau_c}{36} \: T^2 ~~{\rm for~BE}
\nn\\
&=& \frac{ge^2\:\tau_c}{3\pi^2}\zeta(2) \: T^2=\frac{ge^2\:\tau_c}{18} \: T^2~~{\rm for~FD}~,
\label{sig_anl}
\eea
where 
\bea
\zeta(2)&=&\frac{1}{\Gamma(2)}\int_0^\infty x^3/(e^x-1)
\nn\\
&=&\sum_{n=1}^\infty \frac{1}{n^2}=\frac{\pi^2}{6}~.
\eea
From Eqs.~(\ref{eta_anl}) and ~(\ref{sig_anl}), we can understand that
$\eta/(\tau_cT^4)$ and $\sigma/(\tau_c T^2)$ are constant values, whose
magnitude are different for different distribution function as shown by
straight horizontal lines (solid line for MB, dash line for BE and dotted line for FD)
in Fig~\ref{fig:estB_T}(a), (b). Here we did not take any degeneracy factor, i.e. we keep $g=1$.
The constant value of $\eta/(\tau_cT^4)$ and $\sigma/(\tau_c T^2)$ means that shear viscosity
and electrical conductivity of massless bosonic or fermionic matter are proportional to fourth power and second
power of temperature respectively. Another thing is that both transport coefficients $\eta$ and $\sigma$
are proportional to relaxation time in $B=0$ picture, which will be modified at $B\neq 0$ case. We will
see it in next subsection.

During generation of figures in next sub-section (\ref{Shel_Bn0}), 
we did not consider spin, flavor degeneracy as well as 
particle anti-particle contribution. Although, for QGP system where particle anti-particle both are 
present, one must consider these degeneracy factors.
Another point - for zero chemical potential, Hall components of 
shear viscosity ($\eta_{\times}$) and electrical conductivity ($\sigma_{\times}$) will vanish because Hall transport for equal number of particles and anti-particles
are exactly canceled out. One should identify 
that Hall coefficients are odd function of $\tau_B$, which is basically responsible for cancel ling the contributions between particle and anti-particle. At finite chemical potential ($\mu$) hall components will be non-zero. In next 
sub-section (\ref{Shel_Bn0}), results are for 
only one species with $g=1$ and $\mu=0$ to focus only on each components of the transport 
coefficients in a magnetic field. However, in Sec.~(\ref{KSS_Bn0}), instead of one species, 
we have considered entire plasma (massless QGP) at $\mu=0$ with appropriate degeneracy factors, 
where we don't get any Hall transport coefficients. 

\subsection{Shear viscosity and electrical conductivity for $B\neq0$}
\label{Shel_Bn0}
Let us come to $B\neq 0$ picture and applying massless case in the expressions of perpendicular and Hall-type shear viscosity and electrical conductivity components, 
given in Eqs.~(\ref{eta4_B}) or (\ref{eta4_BR}) and (\ref{cond_n}).
Using massless relation $\vp=\om$ in Eqs.~(\ref{eta4_B}) or (\ref{eta4_BR}), one can obtain $\eta_{\perp,\parallel,\times}$ as
\bea
&&\eta_{\perp,\parallel} = {\tilde\eta}_{1,2} =\eta_0, (\eta_0 +\eta_2)
\nn\\
&=&\frac{\eta(B=0)}{1+(4,1)\times(\tau_c/\tau_B)^2}=\frac{\frac{4g\:\tau_c}{5\pi^2} \: T^4}{1+(4,1)\times(\tau_c/\tau_B)^2}
~~{\rm for~MB}
\nn\\
&=& \frac{\eta(B=0)}{1+(4,1)\times(\tau_c/\tau_B)^2}=\frac{\frac{4g\pi^2\:\tau_c}{450} \: T^4}{1+(4,1)\times(\tau_c/\tau_B)^2} ~~{\rm for~BE}
\nn\\
&=& \frac{\eta(B=0)}{1+(4,1)\times(\tau_c/\tau_B)^2}=\frac{\frac{7g\pi^2\:\tau_c}{900} \: T^4}{1+(4,1)\times(\tau_c/\tau_B)^2}~~{\rm for~FD}
\nn\\
\label{eta_2}
\eea
and
\bea
\eta_{\times} &=& {\tilde\eta}_4 =\eta_4
\nn\\
&=&\frac{\eta(B=0)(\tau_c/\tau_B)}{1+(\tau_c/\tau_B)^2}=\frac{\frac{4g\:\tau_c}{5\pi^2} \: (\tau_c/\tau_B)T^4}{1+(\tau_c/\tau_B)^2}
~~{\rm for~MB}
\nn\\
&=& \frac{\eta(B=0)(\tau_c/\tau_B)}{1+(\tau_c/\tau_B)^2}=\frac{\frac{4g\pi^2\:\tau_c}{450} \: (\tau_c/\tau_B)T^4}{1+(\tau_c/\tau_B)^2} ~~{\rm for~BE}
\nn\\
&=& \frac{\eta(B=0)(\tau_c/\tau_B)}{1+(\tau_c/\tau_B)^2}=\frac{\frac{7g\pi^2\:\tau_c}{900} \: (\tau_c/\tau_B)T^4}{1+(\tau_c/\tau_B)^2}~~{\rm for~FD}~.
\nn\\
\label{eta_4}
\eea
For calculation simplification, we have considered average energy in $\tau_B$ (See  Sec.~(\ref{Ap_Eav}) in Appendix)
\bea
\tau_B&=&\om_{\rm av}/eB=\frac{3T}{eB} ~~{\rm for~MB}
\nn\\
&=&\Big\{\frac{\zeta(4)}{\zeta(3)}\Big\}\frac{3T}{eB} ~~{\rm for~BE}
\nn\\
&=&\Big\{\frac{7\zeta(4)}{2\zeta(3)}\Big\}\frac{3T}{eB} ~~{\rm for~FD}~.
\label{Eav_TB}
\eea 

Solving Eq.~(\ref{cond_n}) for massless case, we get
\bea
\sigma_n &=&\frac{\sigma(B=0)(\tau_c/\tau_B)^n}{1+(\tau_c/\tau_B)^2}=\frac{\frac{ge^2\:\tau_c}{3\pi^2}
\: T^2(\tau_c/\tau_B)^n}{1+(\tau_c/\tau_B)^2}~~{\rm for~MB}
\nn\\
&=& \frac{\sigma(B=0)(\tau_c/\tau_B)^n}{1+(\tau_c/\tau_B)^2}=\frac{\frac{ge^2\:\tau_c}{36}
\: T^2(\tau_c/\tau_B)^n}{1+(\tau_c/\tau_B)^2}~~{\rm for~BE}
\nn\\
&=& \frac{\sigma(B=0)(\tau_c/\tau_B)^n}{1+(\tau_c/\tau_B)^2}=\frac{\frac{ge^2\:\tau_c}{18} 
\: T^2(\tau_c/\tau_B)^n}{1+(\tau_c/\tau_B)^2}~~{\rm for~FD}~,
\nn\\
\label{sig_Bm0}
\eea
where reader should keep in mind - $\sigma_{\perp}=\sigma_0$
and $\sigma_{\times}=\sigma_1$.

We have plotted $\eta_{\perp,\parallel,\times}/(\tau_cT^4)$
as a function of $T$ and $eB/m_\pi^2$ in Fig.~\ref{fig:estB_T}(a) and \ref{fig:estB_B}(a).
In similar way, by putting $n=0,1$ in Eq.~(\ref{cond_n}), we can estimate $\sigma_{\perp,\times}/(\tau_cT^4)$,
whose $T$ and $eB/m_\pi^2$ dependence are shown in Fig.~\ref{fig:estB_T}(b) and \ref{fig:estB_B}(b).
The $\eta_{\perp,\parallel}$ and $\sigma_{\perp}$
increases with $T$ and decreases with $B$ due to increasing and decreasing of
anisotropic factor $1/[1+(4,1)\times(\tau_c/\tau_B)^2]$~,
where $\tau_B\propto T/B$. However, this monotonic
trend can not be obtained for Hall-type components of shear viscosity $\eta_{\times}$ and electrical
conductivity $\sigma_{\times}$ because their anisotropic factor 
\be
A_{\times}=(\tau_c/\tau_B)/[1+(\tau_c/\tau_B)^2]
\ee
follow a non-monotonic $T$, $B$ dependence. For constant value of $\tau_c$, $\frac{\tau_c}{\tau_B}\propto\frac{B}{T}$ and so, the anisotropic factor $A_{\times}$
will increase first in the domain of $\frac{\tau_c}{\tau_B}<1$ and then decrease in the 
domain of $\frac{\tau_c}{\tau_B}>1$. Those domains can be seen in Figs.~\ref{fig:estB_T}(c)
and \ref{fig:estB_B}(c), where $\tau_B$ is plotted against $T$ and $B$ axes and compared
with $\tau_c=2$ fm value. One should notice that
the parallel component of electric conductivity
is exactly same with isotropic value i.e. $\sigma_{\parallel}=\sigma$ but for shear viscosity,
they are not equal, rather $\eta_{\parallel}<\eta$.

\begin{figure}
 \resizebox{0.45\textwidth}{!}{
 	\includegraphics{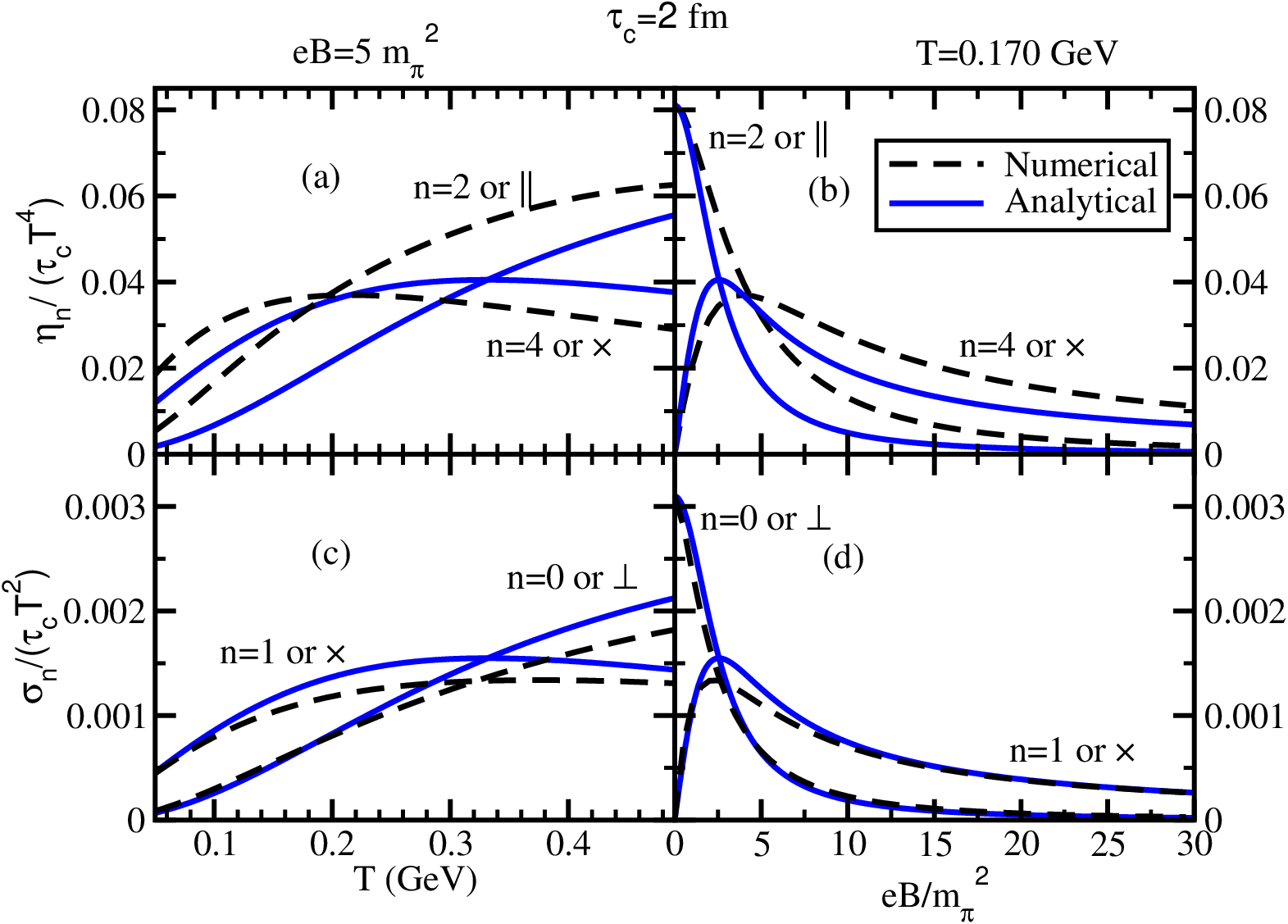}
 }
 \resizebox{0.45\textwidth}{!}{
 	\includegraphics{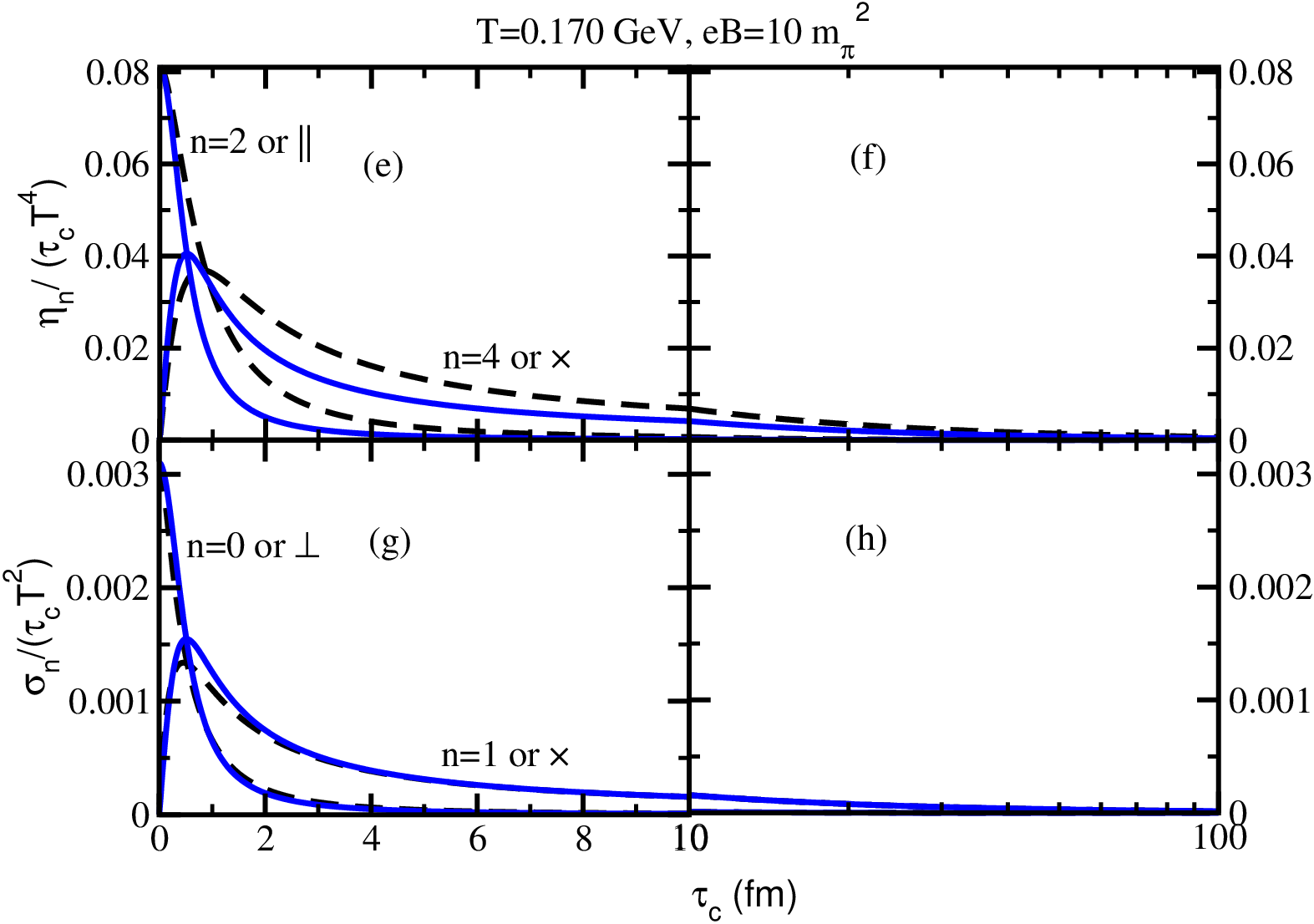}
 }	
 	\caption{Numerical values of $\eta_{2,4}/(\tau_cT^4)$ from Eqs.~(\ref{eta2_B}), (\ref{eta4_B}),
 	and $\sigma_{0,1}/(\tau_cT^4)$ from Eq.~(\ref{cond_n}) are plotted by black dash line. Corresponding
 	values from analytic Eqs.~(\ref{eta_2}), (\ref{eta_4}) and (\ref{sig_Bm0}) are plotted by blue solid line.}
 	\label{fig:Num_Ana}
 \end{figure}
To get a simplified analytic form of $\eta_{\perp,\parallel,\times}$ and $\sigma_{\perp,\times}$, as given in
Eqs.~(\ref{eta_2}), (\ref{eta_4}) and (\ref{sig_Bm0},)
we have considered momentum/energy independent expression 
of $\tau_B=\om_{\rm av}/(eB)$, given in Eq.~(\ref{Eav_TB}).
However, one can obtain numerical values of $\eta_{\perp,\parallel,\times}$ from Eqs.~(\ref{eta4_B}) or (\ref{eta4_BR}),
and $\sigma_{\perp,\parallel,\times}$ from Eq.~(\ref{cond_n}) by using massless relation $\vp=\om$ and $\tau_B=\om/(eB)$,
instead of any simplified average value.
These numerical and analytical results are plotted by black dash and blue solid lines
in Fig.~(\ref{fig:Num_Ana}), where their values are quite separated but well merged
at low $T$ and/or high $B$ and $\tau_c$. Their merging zone is basically strong field
domain, where $\tau_c/\tau_B>>1$ will be achieved. 
Qualitative $T$, $B$ and $\tau_c$ dependent pattern of
curves, obtained from analytic and numerical approaches are quite similar. For simplicity and getting a quick estimations, one can use analytic expressions of massless
case, given by Eqs.~(\ref{eta_2}), (\ref{eta_4}) and (\ref{sig_Bm0}).

\subsection{Comparison between massless expressions of ${\tilde\eta}_n$ and $\eta_n$}
Let us now proceed for comparison between massless expressions of ${\tilde \eta}_n$ and $\eta_n$. 
There are two different possible set of independent traceless
tensors, prescribed in Ref.~\cite{Landau} and Refs.~\cite{XGH1,XGH2}, from where one can find ${\tilde\eta}_n$ and $\eta_n$ respectively. Now if we analyze the earlier
existing references, the we can find that Refs.\cite{Tuchin,G_NJL_B,JD2,HRGB,HM3,Denicol,Asutosh} have adopted former set of tensors and calculated ${\tilde\eta}_n$, While Ref.~\cite{Denicol,BAMPS} have adopted latter set of tensors and calculated $\eta_n$. In this context, present article has obtained both ${\tilde\eta}_n$ and $\eta_n$ in RTA based kinetic theory approach and present section is aimed to explore the comparative documentation
of their massless expressions. Though final expressions of
${\tilde\eta}_n$, obtained/used by Refs.\cite{Tuchin,G_NJL_B,JD2,HRGB,HM3,Asutosh} and $\eta_n$, obtained/used by Refs.~\cite{Denicol,BAMPS} are exactly same as obtained here, but difference in approaches or mathematical steps can be found. For example, to calculate $\eta_n$ Ref.~\cite{Denicol} has
gone through method of moment technique, while present work has adopted RTA based kinetic theory. On the other hand, to calculate ${\tilde\eta}_n$, Refs.~\cite{Asutosh,HRGB} and present work both have adopted
RTA based kinetic theory but one should notice the differences among their way
of calculations.

To get a clear comparative picture between the expressions of ${\tilde\eta}_n$ and ${\eta}_n$, let us first write down
their simplified (massless) structures in Table~(\ref{tab:table1}). 
\begin{table}[ht]
	\begin{center}
	\caption{Taking $\tau_c/\tau_B=x$, simplified structures of shear viscosity components,
	obtained from independent tensors of Ref.~\cite{Landau} (first column) and Refs.~\cite{XGH1,XGH2} (second column) are tabulated in respectively.}
	\label{tab:table1}
	\begin{tabular}{ |c|c|} 
		\hline
		From tensors of Ref.~\cite{Landau} & From tensors of Refs.~\cite{XGH1,XGH2} \\
		 \hline
		${\tilde \eta}_0=\eta$ & $\eta_0=\eta\frac{1}{1+4x^2}$ \\ 
		${\tilde \eta}_1=\eta\frac{1}{1+4x^2}$ &
		${\eta}_1=\eta\Big(\frac{16}{3}\Big)\frac{x^2}{1+4x^2}$ \\ 
		${\tilde \eta}_2=\eta\frac{1}{1+x^2}$ &
		${\eta}_2=\eta\frac{3x^2}{(1+4x^2)(1+x^2)}$ \\
		${\tilde \eta}_3=\eta\frac{x}{1/2+2x^2}$ &
		${\eta}_3=\eta\frac{x}{1+4x^2}$ \\
		${\tilde \eta}_4=\eta\frac{x}{1+x^2}$ &
		${\eta}_3=\eta\frac{x}{1+x^2}$  \\
		\hline
	\end{tabular}
\end{center}
\end{table}
They are expressed in terms of $x=\tau_c/\tau_B$.
The quantity $\xi_B=\tau_c(qB/T)$ of Ref.~\cite{Denicol} and our defined quantity 
$x=\tau_c/\tau_B=\tau_c(qB/3T)$ for MB distribution are connected as $x=\xi_B/3$. Adjusting this proper replacement, one can find that our massless expressions of $\eta_n$ are exactly same as obtained by Ref.~\cite{Denicol}, which are listed in second column of Tab.~(\ref{tab:table1}).
Another point is that Ref.~\cite{Denicol} has taken 
$\eta=\frac{4}{3}\tau_c P=\frac{4g\tau_c}{3\pi^2}T^4$, which is $5/3$ times larger than Eq.~(\ref{eta_anl}),
used here for MB distribution. The famous $1/15$ factor in $\eta$, which is coming due to
the identity of average momenta, given in Eq.~(\ref{k_av}),
which is not considered in Ref.~\cite{Denicol}. Instead of $1/15$, they have considered $1/3$
factor as a three dimensional space average. Although whatever the values of $\eta$, one
can normalize it from the expressions $\eta_n$ and the anisotropic factors
will be our matter of interest. 

The connection between ${\tilde\eta}_n$ and $\eta_n$
are well discussed in Sec.~(\ref{eta_R}). One can easily
check those relation with the help of these massless expression, given in Table.~(\ref{tab:table1}).

\section{Aspects of RHIC or LHC phenomenology}
\label{KSS_Bn0}
\begin{figure}
\resizebox{0.45\textwidth}{!}{
\includegraphics{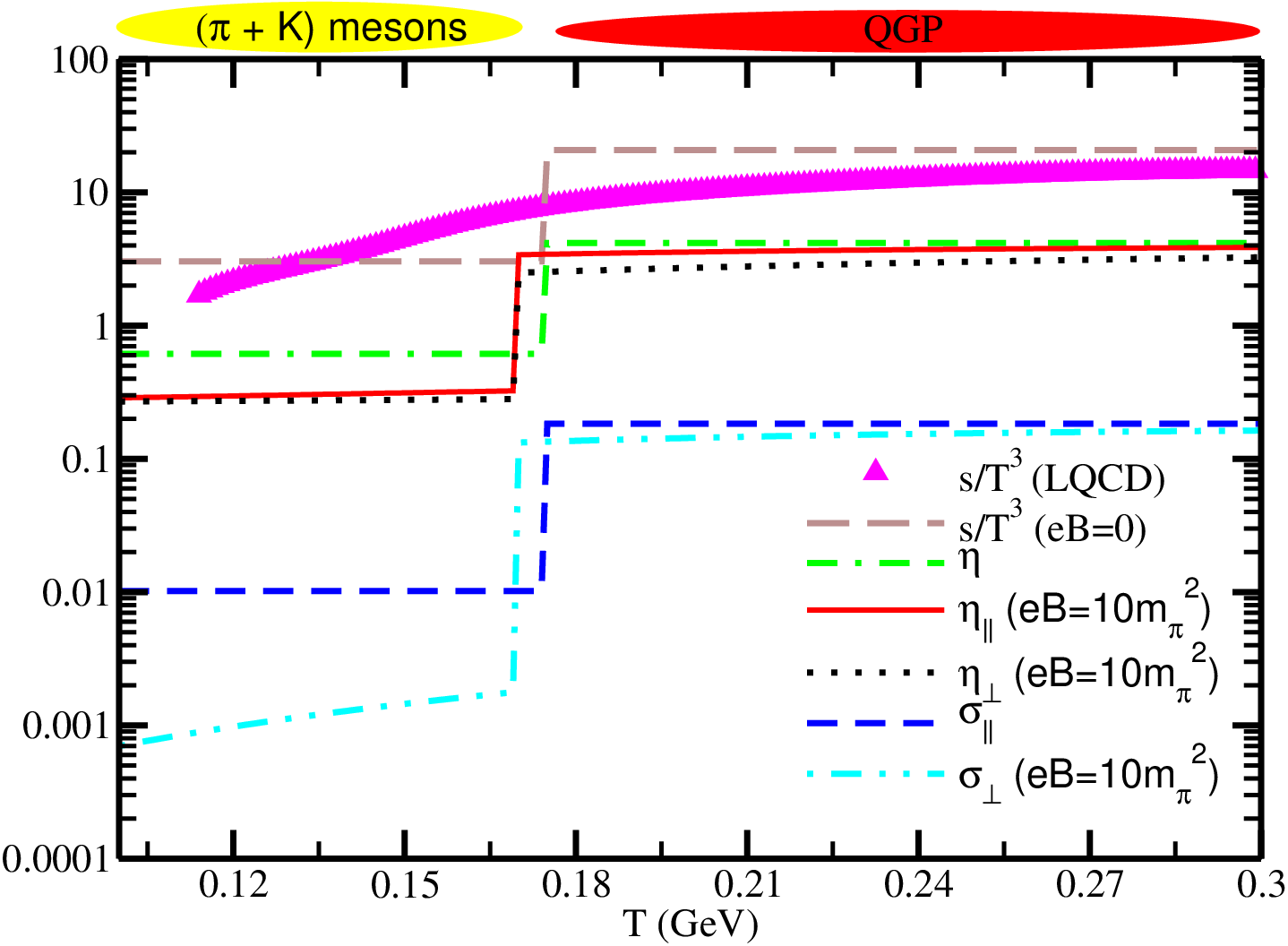}
}
\caption{Normalized values of entropy density $(s/T^3)$, parallel and perpendicular component of shear viscosity ($\eta_{\parallel,\perp}/(\tau_c T^4)$) and electrical conductivity ($\sigma_{\parallel,\perp}/(\tau_c T^2)$) at $B=0$ and $B=10 m_\pi^2$ for HM ($T<T_c$) and QGP ($T>T_c$) massless phases, where transition temperatures $T_c$ are taken from Refs~\cite{Bali1,Bali2}.}
\label{fig:ses_T}
\end{figure}
The mathematical anatomy of shear viscosity and electrical conductivity
in presence of magnetic field, which is addressed for MB, FD and BE distribution
functions in general, can have a straight forward application in RHIC or LHC matter.
The massless results can provide us a rough boundary within which RHIC or LHC matter
undergoes a phase transition from quark gluon plasma (QGP) state to hadronic matter (HM) state.
Here we have attempted to sketch those possible boundaries of two phases in 
Fig.~(\ref{fig:ses_T}). Before going to transport coefficients results, we have first demonstrate
the standard entropy density for calibrating our phenomenological attempt.
Above transition temperature ($T>T_c$), we have considered massless QGP with 
(charged) quark degeneracy factors 
\bea
g&=& {\rm Color}\times{\rm Spin}\times {\rm Particle-Antiparticle}\times {\rm Flavor}
\nn\\
&=&3\times 2\times 2\times 3
\nn\\
&=&36~,
\eea
(neutral) gluon degeneracy factors
\bea
g&=& {\rm Color}\times{\rm Spin}
\nn\\
&=&8\times 2
\nn\\
&=&16~.
\eea
On the other hand, below transition temperature ($T< T_c$), we have considered massless HM, where abundant pion and Kaon are considered only. For $B=0$ case,
we will consider hadron degeneracy factor $g=7$ by counting pion degeneracy factors $g=3$ and Kaon degeneracy factors $g=4$.
while for $B\neq 0$, we have to count degeneracy factors of charged hadrons and neutral
hadrons as $g=4$ ($\pi^+$, $\pi^-$, $K^+$, $K^-$) and $g=3$ ($\pi^0$, $K^0$, ${\bar K}^0$) respectively. 
We will use FD distribution results for quarks and BE distribution results for gluon and mesons.
Taking care about those distributions and degeneracy factors of two phases, we have included the standard
massless results of normalized entropy density $s/T^3$ (brown dash line) along with its lattice QCD (LQCD) results (pink triangles)~\cite{Bali1,Bali2} at $B=0$. The crossover
nature of quark-hadron transition, prescribed by LQCD can be realized as a smooth
transition between massless limits of two phases. So, these massless estimations might
be considered as reference frames of two phases within which actual crossover transition
will be occurred.

After getting calibration from $s/T^3$ results, let us proceed to sketch the boundaries
of two phases for normalized shear viscosity ($\eta/(\tau_c T^4)$), electrical conductivity
($\sigma/(\tau_c T^2)$). At $B=0$, they are drawn by green dash-dotted and blue dash lines,
whose shape are exactly same as $s/T^3$ but their strengths are different only.
Marking the values of their boundaries, we can write the inequalities $3.04<\frac{s}{T^3}<20.79$,
$0.61<\frac{\eta}{\tau_c T^4}<4.16$, $0.01<\frac{\sigma}{\tau_c T^2}<0.18$ at $B=0$
for any values of $\tau_c$. Now when we go for $B\neq 0$
case of QGP and HM, then neutral hadrons and gluons will
have same contribution in shear viscosity as they had for $B=0$ case. However, the contribution from charged hadrons and quarks will face a suppression due to magnetic field. The net suppression values of $\eta_\perp$ and $\eta_\parallel$ are shown in Fig.~(\ref{fig:ses_T}) by red solid and black dotted lines respectively. The results of $\eta_\perp\neq \eta_\parallel$ indicates that anisotropy in shear transportation is developed due to magnetic field.
Similarly, when we go to calculate electrical conductivity, we will find suppression of contribution
from charged hadron and quarks. Unlike to shear viscosity case, here neutral hadrons and gluon will never contribute. Another difference from shear viscosity case,
at finite $B$ picture $\sigma_\perp$ only suppress but $\sigma_{\parallel}$ remain equal with $\sigma$. So, the differences between $\sigma_\perp$ and $\sigma_{\parallel}$ in QGP and HM phases basically measure the anisotropic trend of electric charge transportation. The impact of this anisotropic conductivity might disclose an anisotropic dilepton
and photon productions, as one can find a proportional
relation in Ref.~\cite{Nicola}. The detail four or three momentum distribution of electromagnetic current-current corelator is basically hidden in dilepton or photons yields, whose zero momentum limit measures the electrical conductivity. That is why a straight forward
expectation is a difference between dilepton or photon yields in parallel and perpendicular direction as $\sigma_{\parallel}\neq\sigma_{\perp}$. Its detail derivations are quite rigorous, so we plan it as a
future project.

\begin{figure}
\resizebox{0.45\textwidth}{!}{
\includegraphics{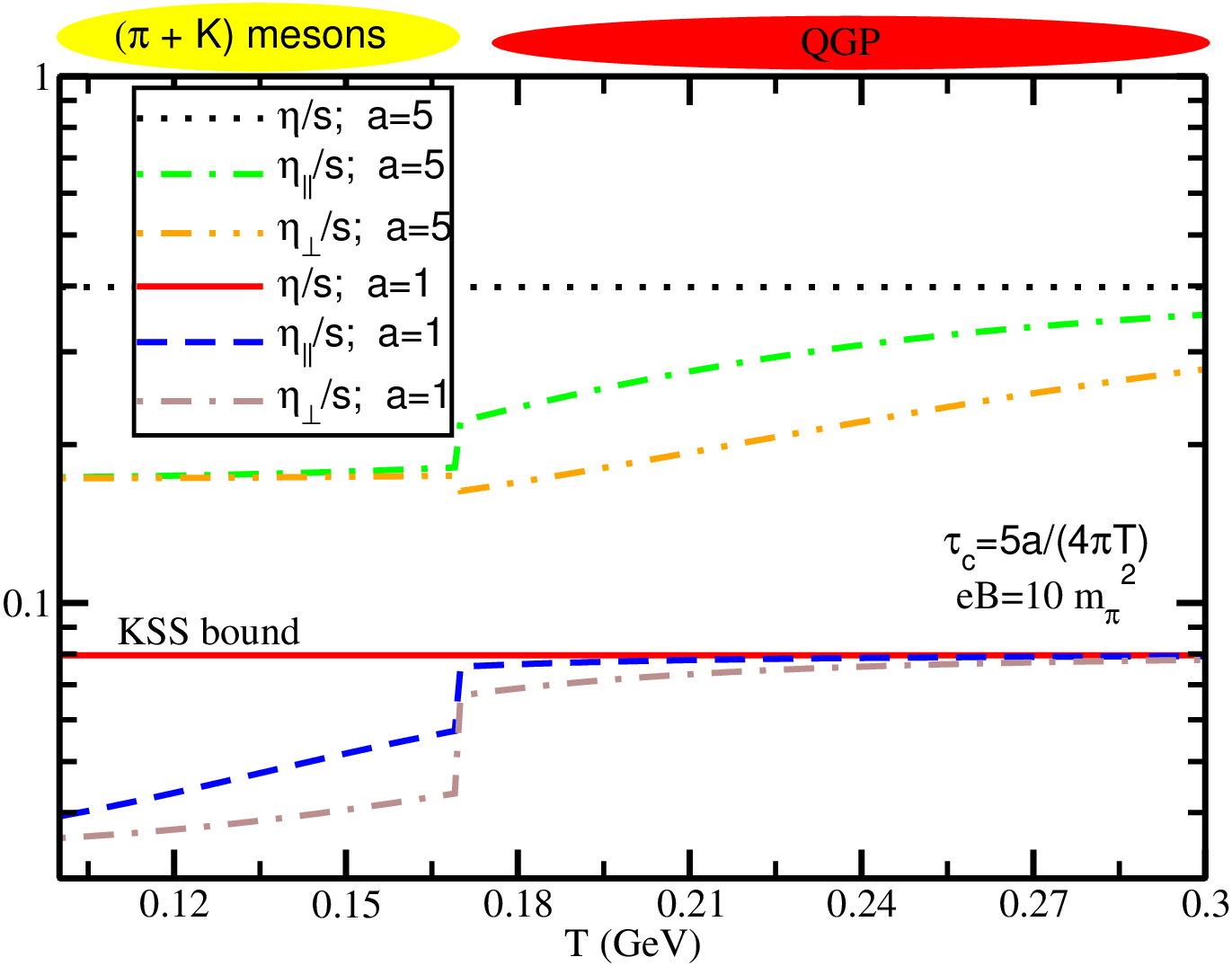}
}
\caption{Numerical bands for viscosity to entropy density ratio at $B=0$ and $B=10 m_\pi^2$.}
\label{fig:eta2bys}
\end{figure}
We have not discussed anything about Hall component $\eta_{\times}$
and $\sigma_{\times}$ as they are completely vanished for QGP system
with zero quark chemical potential and HM system with zero baryon
chemical potential. This is happened due to opposite Hall-flow of particle and anti-particle with equal strength at zero net quark/baryon density system. However, for non-zero net quark/baryon density system,
one can get a non-zero Hall flow, but we have not gone through that direction as our interest is focus on RHIC or LHC phenomenology.

From the massless values of thermodynamical quantities like entropy density
and transport coefficients like shear viscosity, electrical conductivity for QGP and 
HM phases, we might get a range, within which their actual values will change. From lattice QCD (LQCD) side~\cite{Bali1,Bali2}, quark-hadron phase transition is realized
as crossover type transition, which shows inverse magnetic catalysis (IMC) pattern near transition temperature. This fact is attempted to map from effective QCD model side~\cite{IMC1,IMC2,IMC3} (see Ref.~\cite{IMC1} for review). With the help of those effective
QCD model, one can get rich version of estimations for $\eta_{\parallel,\perp}$, $\sigma_{\parallel,\perp}$ within the massless boundaries of two phases.

From QGP phenomenology, we know that 
the viscosity to entropy density ratio ($\eta/s$)
is more meaningful quantity than viscosity only, as
the ratio directly measures the fluid property.
RHIC or LHC matter has very low $\eta/s$, close
to its KSS bound - $\frac{1}{4\pi}$~\cite{KSS}.
To get experimental values of $\frac{\eta}{s}=\frac{a}{4\pi}$ (with $a=1-5$)~\cite{Exp_Rev} for RHIC or LHC matter, 
$\tau_c=\frac{5a}{4\pi T}$ is expected for massless fluid, by using $\eta$ from Eq.~(\ref{eta_anl}) 
and $s$ from Eq.~(\ref{s_anl}). Interestingly $\tau_c$ are same for MB, BE and FD cases, 
although their individual $\eta$ and $s$ expressions are different. 
For QGP case, we have to use the results of BE for gluon and FD for quarks with their
respective degeneracy factors, which will be ultimately canceled out in ratio $\eta/s$.
Hence, the phenomenological imposition to massless QGP,
\bea
\frac{\eta_{\rm QGP}}{s_{\rm QGP}}&=&\frac{a}{4\pi}
\nn\\
\frac{[\frac{7}{8}\times24+16]\frac{4}{5\pi^2}\tau_c\zeta(4)T^4}
{[\frac{7}{8}\times 24+16]\frac{4}{\pi^2}\zeta(4)T^3}&=&\frac{a}{4\pi}
\nn\\
\Rightarrow \tau_c&=&\frac{5a}{4\pi T}~,
\eea
will provide us a numerical band of relaxation time $\tau_c=\frac{5}{4\pi T}$ to
$\frac{25}{4 \pi T}$. If we consider same calculation for massless $(\pi+K)$ system, results remain same because two massless phases are just different by their degeneracy factors and they are exactly canceled out in 
$\eta/s$ ratio. Therefore, two horizontal lines $\eta/s=1/(4\pi)$ and $\eta/s=5/(4\pi)$ at $B=0$ are extended from QGP to HM temperature domain. However, this exact cancellation will not work for $B\neq 0$ case as shear viscosity expressions of charged and neutral particles of QGP and HM system will be different.
Using this $\tau_c$ we have generated $\eta_{\parallel}/s$ and $\eta_{\perp}/s$ for QGP and HM system (in their respective $T$-zones) in 
Fig.~(\ref{fig:eta2bys}), where blue dash, brown dash-dotted
 lines correspond to $\tau_c=\frac{5}{4\pi T}$ and green dash-dotted, orange dash-double-dotted lines correspond to $\tau_c=\frac{25}{4 \pi T}$. We notice that
the horizontal lines (red solid and black dotted lines) for $\eta/s=1/(4\pi)$ and $5/(4\pi)$, indicating numerical band
for $\eta/s$ at $B=0$, are suppressed in finite $B$ scenario. We can understand that
the suppression is mainly coming for $B$ dependent shear viscosity of quark component in QGP temperature range ($T>T_c$) and charged hadron component in HM temperature range ($T<T_c$). Although for realistic quark-hadron
crossover type transition might have a more reach $T$
and $B$ dependent structure but these massless values of two phases provide us a rough order of magnitude and a qualitative conclusion about the impact of $B$ on fluid property of QGP and HM system. The present work indicates that the magnetic field might be in favor to
build perfect fluid nature of RHIC or LHC matter as
the ratio is suppressing at finite magnetic field picture.

\section{On lower bound of viscosity to entropy density ratio}
\label{sec:KSS}
If we notice the curves of $\eta_\parallel/s$ and $\eta_\perp/s$ in 
Fig.~(\ref{fig:eta2bys}) at $B\neq 0$ for $\tau_c=5/(4\pi T)=\tau_c^0$ (say), then we can see that they come lower than KSS bound. Now according to dual holographic type theory~\cite{ADS1,ADS2}, parallel component can cross the bound but perpendicular can't i.e. their lower bound can be expressed as
\bea
\frac{\eta_{\parallel}}{s} &<& \frac{1}{4\pi}
\nn\\
\frac{\eta_{\perp}}{s} &=& \frac{1}{4\pi}~.
\eea
If we analyze the existing litterateurs on 
QGP~\cite{Tuchin,G_NJL_B,Asutosh,HRGB,Denicol,BAMPS} including present work,  we can't reach lower bound
$\frac{\eta_{\perp}}{s} = \frac{1}{4\pi}$ by using
$\tau_c$. The reason might be as follows. The quantum description like Landau quantization version of kinetic theory approach or Kubo approach has not been considered
yet. They might help to build that bound probably but it will be only confirmed if one will really enter into those calculations in future. However, one can always able to find a $\tau_c(T,B)$, for which $\frac{\eta_{\perp}}{s} = \frac{1}{4\pi}$ will be established. 
\begin{figure}
\resizebox{0.45\textwidth}{!}{
\includegraphics{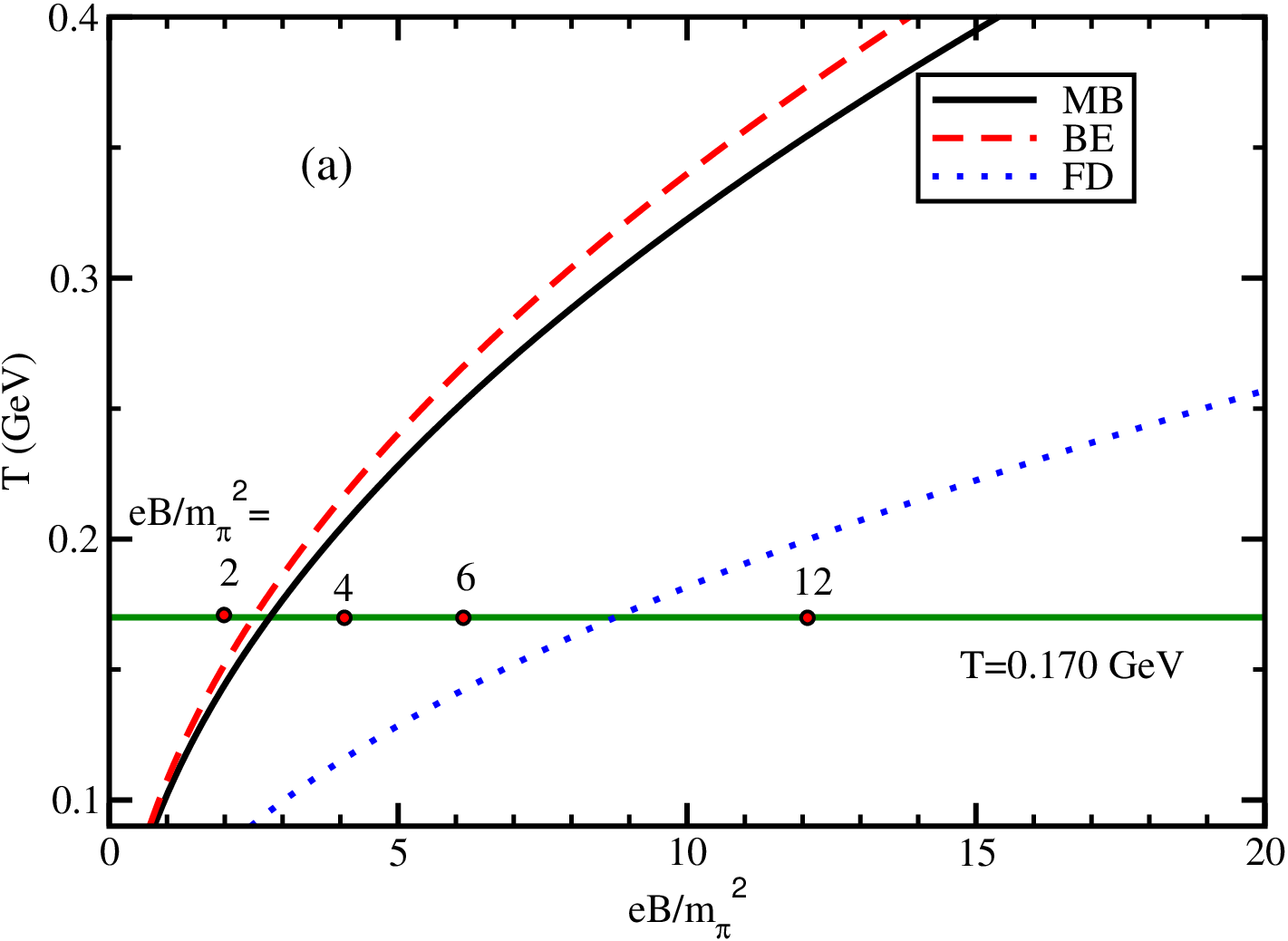}
}
\caption{$T\propto\sqrt{eB}$ curves for MB (black solid line), BE (red dash line), 
FD (blue dotted line) cases, below which $\eta_{\perp}/s$ never touch KSS line.}
\label{fig:TB_KSS1}
\end{figure}
\begin{figure}
\resizebox{0.45\textwidth}{!}{
\includegraphics{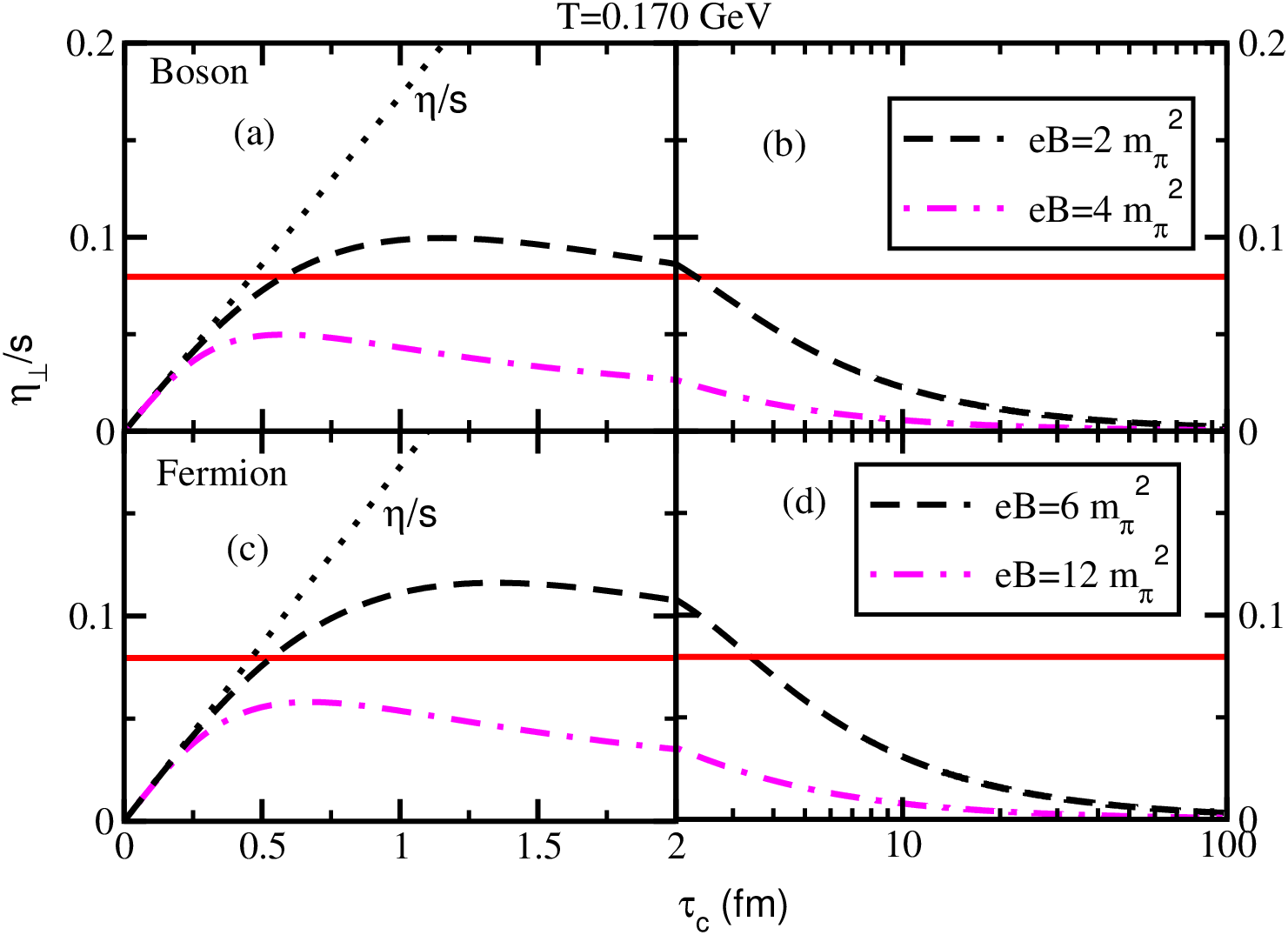}
}
\caption{$\eta_{\perp}/s$ vs $\tau_c$ for bosonic (a-b) and fermionic (c-d) medium. Black dotted line is for $B=0$ case; black dash and pink dash-dotted lines are for two finite magnetic field cases, where pink dash-dotted lines always remain below KSS bound.}
\label{fig:TB_KSS2}
\end{figure}
Similar to $B=0$ case, where $\tau_c^0=5/(4\pi T)$ is
obtained by imposing $\frac{\eta}{s}=\frac{1}{4\pi}$ to massless MB/BE/FD system, we can impose
\bea
\frac{\eta_{\perp}}{s}&=&\frac{1}{4\pi}
\nn\\
\Rightarrow \frac{T\tau_c/5}{1+4(\tau_c/\tau_B)^2}&=&\frac{1}{4\pi}
\nn\\
\Rightarrow \frac{\tau_c/\tau_c^0}{1+4(\tau_c/\tau_B)^2}&=& 1~,
\label{es_KSSB}
\eea
where average $\tau_B$ values for MB, BE and FD cases
will be different.
Above Eq.~(\ref{es_KSSB}) will provide us a quadratic equation of $\tau_c$:
\be
\tau_c^2 - \Big(\frac{ \tau_B^2}{4\tau^0_c}\Big)\tau_c+\frac{\tau_B^2}{4} = 0~,
\ee
whose solution is
\be
\tau_c=\tau_c^{\pm}=\frac{\tau_B^2}{8\tau^0_c}\Big[1\pm\sqrt{1-16\Big(\frac{\tau^0_c}{\tau_B}\Big)^2}\Big]~.
\label{tauc2}
\ee
So far from our best knowledge, we are first time addressing an {\it analytic} expressions of $\tau_c(T,B)$,
where massless bosonic/fermionic matter in presence of magnetic field reach the KSS bound.
To get a physical solution of Eq.~(\ref{tauc2}), we need
\bea
1-16\Big(\frac{\tau^0_c}{\tau_B}\Big)^2 &\geq &0
\nn\\
\Rightarrow \tau_B &\geq & 4\tau_c^0~.
\eea
Using MB relation $\tau_B=\frac{3T}{eB}$ in above inequality, we have
\bea
\frac{3T}{eB} &\geq & 4\frac{5}{4\pi T}
\nn\\
T &\geq & \Big(\frac{5 eB}{3\pi}\Big)^{1/2}~.
\label{MB_TeB}
\eea
Corresponding FD and BE relations from Eq.~(\ref{Eav_TB}) will give 
\bea
T &\geq & \Big[\Big(\frac{\zeta(3)}{\zeta(4)}\Big)\frac{5 eB}{3\pi}\Big]^{1/2}~{\rm for ~BE}
\nn\\
T &\geq & \Big[\Big(\frac{2\zeta(3)}{7\zeta(4)}\Big)\frac{5 eB}{3\pi}\Big]^{1/2}~{\rm for ~FD}~.
\label{FDBE_TeB}
\eea
Drawing $T-eB$ curves of Eqs.~(\ref{MB_TeB}) and (\ref{FDBE_TeB}) in Fig.\ref{fig:TB_KSS1},
one can identify upper allowed domain, where KSS bound can be achieved. On the other hand, lower domain is
forbidden zone, because we will get $\eta_{\perp}/s<1/(4\pi)$, which is not possible according to Refs.~\cite{ADS1,ADS2}. 
 Or in other word, we should accept that the present massless expression of $\eta_{\perp}/s$ fails to maintain its lower bound in the forbidden zones of $T$-$B$ plane which
can also be consider as strong magnetic field zone.

To explore the 
fact, we have drawn straight horizontal (green solid) line at $T=0.170$ GeV, where we have chosen
points $eB=2m_\pi^2$ and $eB=4m_\pi^2$ in allowed and forbidden zone for bosonic medium. Similar
points are $eB=6m_\pi^2$ and $eB=12m_\pi^2$ for fermionic medium.
Generating $\eta_{\perp}/s$ vs $\tau_c$ curves for bosonic medium (a-b) and fermionic (c-d) medium at those
points, we can see that $\eta_{\perp}/s$ always remain below KSS value at the points of forbidden zone.
At the point of allowed zone, we can get solution of Eq.~(\ref{tauc2}) at $\tau_c=\tau_c^{\pm}$,
which can be identified by crossing points of $\eta_{\perp}/s$ with KSS line (red horizontal line) in 
Figs.~(\ref{fig:TB_KSS2}). We have also drawn $\eta/s\propto\tau_c$ (dotted line) curves
in Figs.~\ref{fig:TB_KSS2}(a) and (c). Due to simple proportional nature, $\eta/s$ will cross KSS line
at one point ($\tau_c=\tau_c^0$), but $\eta_{\perp}/s$ cross the KSS line in two points ($\tau_c=\tau_c^{2\pm}$)
because of non-monotonic relation $\eta_{\perp}/s\propto \frac{\tau_c}{1+4(\tau_c/\tau_B)^2}$. We notice that
$\tau_c^0$ and $\tau_c^-$ are little close in numerical values and both signify the lower values of $\tau_c$, for which
viscosity to entropy density ratio touch its KSS bound. Interestingly, we are getting 
an upper value of $\tau_c$ ($\tau_c^+$), where $\eta_{\perp}/s$ again reach its KSS bound. This
fact is completely new fact, appeared in the picture of finite magnetic field.
\begin{figure}
\resizebox{0.45\textwidth}{!}{
\includegraphics{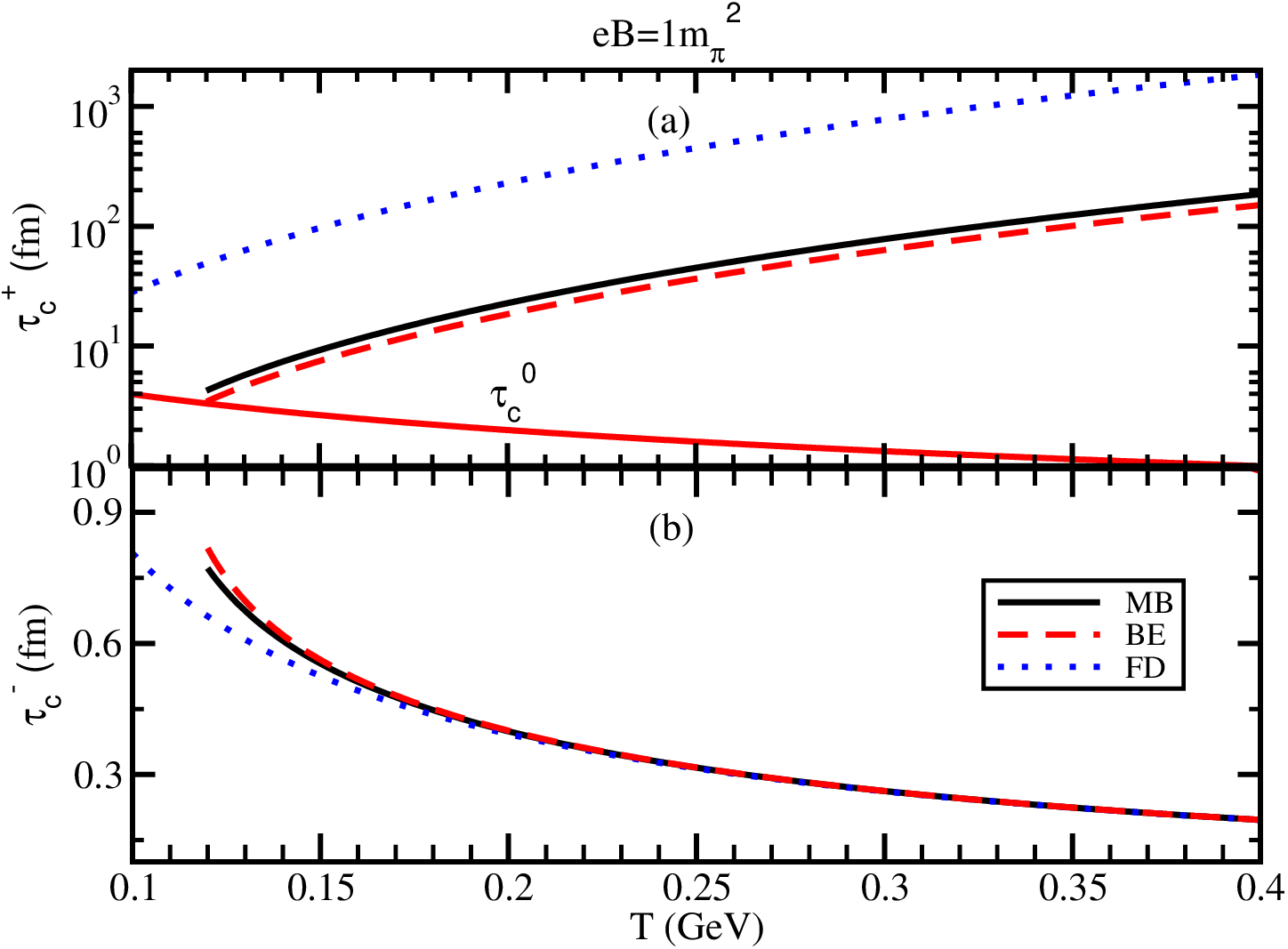}
}
\resizebox{0.45\textwidth}{!}{
\includegraphics{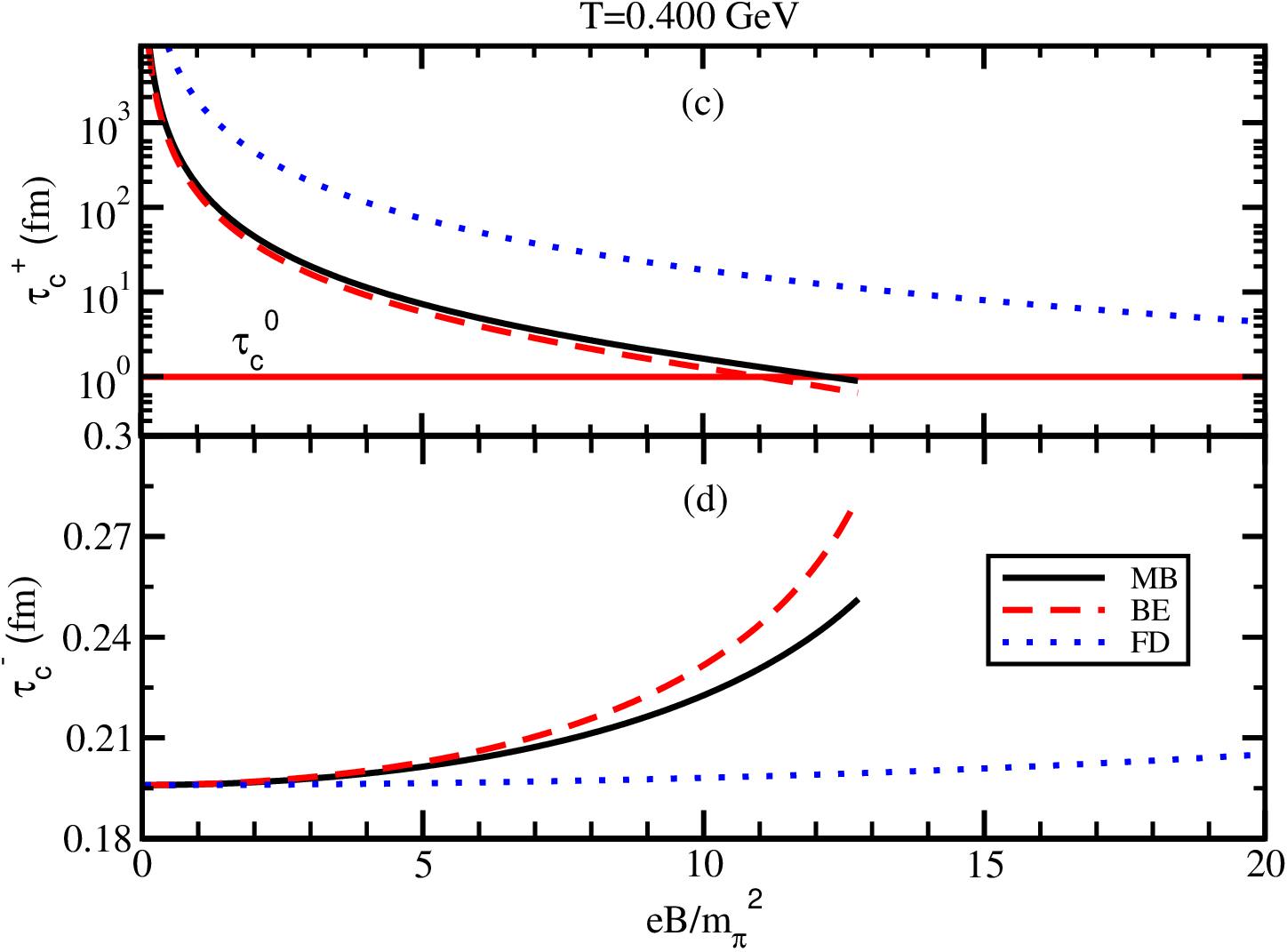}
}
\caption{$T$ (a, b) and $eB/m_\pi^2$ (c, d) dependence of $\tau_c^+$ (a, b), 
	$\tau_c^-$ (c, d) for MB, BE, FD cases.}
\label{fig:tc_KSS_TeB}
\end{figure}
Above graphical discussion give us an idea about allowed/forbidden $T$-$B$ domain,
where $\tau_c$ parameter can/can't tune $\eta_{\perp}/s$ to $1/(4\pi)$. Now, let us 
draw $\tau_c^{\pm}(T,B)$ along with $\tau_c^0(T)$ in Fig.~(\ref{fig:tc_KSS_TeB})
 by using Eqs.~(\ref{tauc2}). We have filtered out 
 unphysical points of $\tau_c^{\pm}(T,B)$. Fig.~\ref{fig:tc_KSS_TeB}(a), (c)
shows $\tau_c^{+}$ increases with $T$ and decreases with $B$, whereas $\tau_c^{-}$ follow
 a mild decrement (increment) with $T$ ($B$) as show in Fig.~\ref{fig:tc_KSS_TeB}(b), (d). Hence, using $\tau_c^{\pm}$, given in Eq.~(\ref{tauc2}), we get lower bound relations
 \bea
 \frac{\eta_{\perp}}{s}&=&\frac{1}{4\pi}
 \nn\\
 \frac{\eta_{\parallel}}{s}&=&
 \frac{\tau_c^{\pm}T/5}{1+\Big(\frac{\tau_c^{\pm}}{\tau_B}\Big)^2}
 \nn\\
 &=&\frac{1}{4\pi}
 \frac{1}{\Big[1+\Big(\frac{\tau_B}{\tau_c^{\pm}}\Big)^2\Big]\Big[1\pm\sqrt{1-16\Big(\frac{\tau_c^0}{\tau_B}\Big)^2}\Big]}
 \nn\\
 &<&\frac{1}{4\pi}~,
 \label{epKSS}
 \eea
whose qualitative dependence are quite agreements
with dual holographic type results~\cite{ADS1,ADS2}~.
 
Now, from fig.~{\ref{fig:TB_KSS1}} it is clear that
our proposed expression $\tau_c^{\pm}(T,B)$ does not have any 
existance in strong field ($T^2/eB \ll 1$) zone or forbidden zone. It might be considered as drawback of our proposed expressions.
Therefore, checking strong field approximation, proposed by Refs.~\cite{ADS1,ADS2},
\be
\frac{\eta_{\parallel}}{s} \approx \pi \frac{T^2}{qB}~,
\ee
is not possible from our proposed picture. However, one can easily
check the opposite limit - weak magnetic field and/or high temperature,
which is allowed by all three types of distributions - MB, BE, FD.
In this limit, $eB/T^2 \ll 1$, i.e. $\tau_c^0/\tau_B \ll 1$ and Eq.~(\ref{tauc2}) can be approximate as,
\bea
\nn
\tau_c^+ &\approx& \frac{\tau_B^2}{4~\tau_c^0}\rightarrow\infty;\\
\tau_c^- &\approx& \tau_c^0~. 
\label{tcKSS}
\eea 
So one can get back our earlier lower bound of $B=0$ picture:
\be
\frac{\eta_{\parallel}}{s}\rightarrow \frac{1}{4\pi}\leftarrow\frac{\eta_{\perp}}{s}~.
\ee

\section{Summary} 
\label{sum}
In summary, we have explored the shear viscosity and electrical conductivity calculations for 
bosonic and fermionic medium, facing an external magnetic field. Unlike to one shear viscosity coefficients in field-free picture,
5 shear viscosity coefficients will come into the
picture, when an external magnetic field is applied to
the medium. This is because 5 independent velocity gradient tensors can be designed in presence of magnetic field and connecting them with the viscous stress tensor, one can get 5 proportional constants, which can be recognized as shear viscosity coefficients (by definition). After going through the existing litterateurs on the relation between viscous stress tensor and 5 velocity gradient tensors at finite magnetic field, we have found two sets of tensor structures. One is proposed from Ref.~\cite{Landau} and 
another is designed by Refs.~\cite{XGH1,XGH2}.

Starting from both sets of tensors one by one, present work has obtained two sets of expressions for 5 shear viscosity coefficients, whose inter-connecting relations are discussed and they have been ultimately classified into three basic components - parallel, perpendicular and Hall components as one get same for electrical conductivity at finite magnetic field. Although, one should keep in mind that former is defined for four rank tensor, while latter is defined for two rank tensor.
We have adopted kinetic theory approach in relaxation time approximation.
For field-free picture, connecting the macroscopic (hydrodynamic) and microscopic (kinetic theory based) structures of viscous stress tensors, one can obtain the shear viscosity expression, where a proportional dependence with relaxation time is established in microscopic structure with the help of RTA based Boltzmann equation. At finite magnetic field,
repeating same mathematical steps, we have obtained 2 sets of 5 shear viscosity components, whose final expressions are in well agreements with earlier references although a difference in methodology or steps can be clearly noticed.

After derivation of expressions, we have presented the parallel, perpendicular and Hall components of shear viscosity and electrical conductivity for massless bosonic and fermionic matter, which will be controlled by
their own distribution functions - Bose-Einstein and Fermi-Dirac. We have also included
the case of Maxwell-Boltzmann distribution functions. In external magnetic
field, a magnetic time scale for cyclotron motion of charge particle in the medium 
is introduced along with relaxation time scale. The competition between these two 
time scale basically create anisotropic transportation in the medium, for which many components are found.
Mainly the values of parallel and perpendicular components will be different at finite magnetic field.
At zero field limits, their values will be merged to their isotropic value.
Taking average energy approximation in magnetic time scale, we get analytic expressions
for all components of shear viscosity and electrical conductivity, which are little deviated from exact numerical values but their qualitative trends remain same. So, for quick estimations, the analytic expressions can be used by the research community.
perpendicular component always reduce with magnetic field and enhance with temperature,
while Hall component follow a non-monotonic trend with temperature and magnetic field.

Using these massless results of viscosity and conductivity of Fermionic and Bosonic medium to
quark gluon plasma and hadronic matter system for 
zero quark/baryon chemical potential, we have obtained
their parallel and perpendicular components, where particle and anti-particle distributions are simply added. Being odd function of electric charge, Hall components will be zero as particle and anti-particle contributions are exactly canceled out. One can get its non-zero values for non-zero quark/baryon chemical potential. The massless results of quark gluon plasma and hadronic matter phases can provide us a rough order of strength, within which actual results will vary during (crossover type) quark-hadron phase transition.
Our results show a suppression of parallel and perpendicular components of shear viscosity as well as viscosities to entropy density ratio due to magnetic field with respect to their isotropic value for zero magnetic field case. It indicates that magnetic field might be in favor of building perfect fluid nature in RHIC or LHC matter. For electrical conductivity, perpendicular component face similar suppression but parallel component remain unaffected.
Also two components have different values, which reflect the anisotropic transportation due to magnetic field and they might have some phenomenological impact, for which a further rigorous research might be necessary. One of the possibilities might be to search
anisotropic dilepton or photon production as they have
direct link with conductivity. We have also attempted to connect our viscosity to entropy density ratio along parallel and perpendicular direction with their lower bounds, expected from dual holographic type theory. Imposition of this connection provide an expressions of relaxation time as a function of temperature and magnetic field, for which lower bound expectation can be achieved.

Our calculation is completely based on classical picture, although we have considered 
Fermi-Dirac and Bose-Einstein distribution, which might be considered as quantum aspect
of statistical mechanics. So we might call this description as semi-classical framework.
However, complete quantum mechanical description by considering Landau levels of charge
particle of medium can be considered as its immediate extension of present framework, which
we are planning for future studies.

{\bf Acknowledgment:} 
JD and SG acknowledge to MHRD funding via IIT Bhilai. PM thanks for (payment-basis)
hospitality from IIT Bhilai during his summer internship tenure (May-June, 2019).

\section{Appendix: $C_n$ calculations}
\label{App1}
\subsection{For tensors of Refs.~\cite{Landau}}
\label{App1_L}
The detail calculations of $C_n$ for traceless independent tensors of Refs.~\cite{Landau} from the RTA based RBE, given in Eq.(\ref{RBE1}) will be documented in this section.

For zero bulk viscosity and knowing that the tensors $V_{ij}$ is symmetric,
$b_{ij}$ is anti-symmetric in $i,j$ we get some relations:
 $\mathbf{\nabla} \cdot \mathbf{V}=V_{ii}=0$, 
$V_{ij}b_{i}b_{j}=0$,
$b_{ij}b_{i}=b_{ij}b_{j}=0$,
$b_{ij}v_{i}v_{j}=0$. 
Using the above conditions, Eq.~(\ref{RBE1}) takes the form,
\bea
\nonumber \frac{\omega}{T} v_{i}v_{j}V_{ij}f_0(1-a f_0) &=& -\frac{2}{\tau_B}\bigg[C_1\Big\{2V_{ik}b_{ij} v_{j} v_{k}\\
\nn
&-& 2V_{ik}b_{ij}b_{k}v_{j} (\vec v \cdot \vec{b})\Big\}\\
\nonumber 
&+& C_2\Big\{2V_{ik}b_{ij}v_{j}b_{k}(\vec v \cdot \vec{b})
\Big\}\\
\nonumber 
&+& C_3\Big\{2V_{ij}v_{i}v_{j} - 4V_{ij}v_{i}b_{j}(\vec v \cdot \vec{b})\Big\}\\
\nonumber 
&+& C_4\Big\{2V_{ij}v_{i}b_{j}(\vec v \cdot \vec{b})\Big\}\bigg]\\
\nonumber 
&+& \frac{1}{\tau_c}\bigg[C_1\Big\{2V_{ij} v_i v_j - 4V_{ij}v_{i}b_{j}(\vec v \cdot \vec{b})\Big\}\\
\nonumber 
&+& C_2\Big\{4V_{ij} v_i b_j(\vec v \cdot \vec{b})\Big\}\\
\nonumber 
&+&C_3\Big\{2V_{ik}b_{jk}v_i v_j - 2 V_{ij}b_{ki}b_{j}v_{k}(\vec v \cdot \vec{b})\Big\}\\
&+& C_4\Big\{4V_{ij}b_{ki}b_{j} v_{k}(\vec v \cdot \vec{b})\Big\}\bigg]
\label{RBEa}
\eea

Now, from eq.~({\ref{RBEa}}) comparing the following tensor structures,

$V_{ij}v_i v_j $ :\\
\be
-\frac{\om}{T}f_0 (1-af_0) = \frac{4}{\tau_B}C_3 - \frac{2}{\tau_c}C_1
\ee
$V_{ij}b_{ik}v_k v_j$ :\\
\be
\frac{4}{\tau_B}C_1 + \frac{2}{\tau_c}C_3 = 0
\ee
$V_{ij}b_{ik}v_k b_j (\vec v \cdot \vec{b})$ :\\
\be
\frac{2}{\tau_B}(-2C_1 + 2C_2) + \frac{1}{\tau_c}(-2C_3 + 4C_4) = 0
\ee
$V_{ij}v_i b_j (\vec v \cdot \vec{b})$ :\\
\be
\frac{2}{\tau_B}(-4C_3 + 2C_4) - \frac{1}{\tau_c}(-4C_1 + 4C_2) = 0
\ee
Solving above four equations we finally get $C$s as,
\bea
\nn
C_1&=&-\frac{\omega}{2T}\;
\frac{\tau_c}{4\left\{\frac{1}{4}+(\tau_c/\tau_B)^2\right\}} ~ f_0(1-a f_0)\\
\nn
C_2&=&-\frac{\omega}{2T}\; \frac{\tau_c}{1+(\frac{\tau_c}{\tau_B})^2} ~ f_0(1-a f_0)\\
\nn
C_3&=&-\frac{\omega}{2T}\; \frac{\tau_c(\frac{\tau_c}{\tau_B})}{2\left\{\frac{1}{4}+(\tau_c/\tau_B)^2\right\}}
~ f_0(1-a f_0)\\
C_4&=&-\frac{\omega}{2T}\; \frac{\tau_c(\frac{\tau_c}{\tau_B})}{1+(\frac{\tau_c}{\tau_B})^2} ~ f_0(1-a f_0)
\label{C'sa}
\eea

\subsection{For tensors of Refs.~\cite{XGH1,XGH2}}
\label{App1_R}
Here, $C_n$'s for tensor of eq.~{\ref{Cmn_R}}{\cite{XGH1,XGH2}} will be calculated.

Using $C_{ij}$s of Eq.~({\ref{Cmn_R}}) in Eq.~({\ref{RBE1}}) with tensor identities 
used in above subsection we get,
\bea
\nonumber \frac{\omega}{T} v_{i}v_{j}V_{ij}&f_0&(1-a f_0) = -\frac{2}{\tau_B}\bigg[C_0\left(2V_{ik}b_{ij} v_{j} v_{k}\right)\\
\nonumber 
&+& C_2\Big\{2V_{ik}b_{ij}v_{j}b_{k}(\vec v \cdot \vec{b})
\Big\}\\
\nonumber 
&+& C_3\Big\{4V_{ij}v_{i}v_{j} - 8V_{ij}v_{i}b_{j}(\vec v \cdot \vec{b})\Big\}\\
\nonumber 
&+& C_4\Big\{2V_{ij}v_{i}b_{j}(\vec v \cdot \vec{b})\Big\}\bigg]\\
\nonumber 
&+& \frac{1}{\tau_c}\bigg[C_0\left(2V_{ij} v_i v_j\right)\\
\nonumber 
&+& C_2\Big\{4V_{ij} v_i b_j(\vec v \cdot \vec{b})\Big\}\\
\nonumber 
&+& C_3\Big\{4V_{ik}b_{jk}v_i v_j - 4 V_{ij}b_{ki}b_{j}v_{k}(\vec v \cdot \vec{b})\Big\}\\
&+& C_4\Big\{4V_{ij}b_{ki}b_{j} v_{k}(\vec v \cdot \vec{b})\Big\}\bigg]~~
\label{RBEb}
\eea

Now, from eq.~({\ref{RBEb}}) comparing the following tensor structures,

$V_{ij}v_i v_j $ :\\
\be
-\frac{\om}{T}f_0 (1-af_0) = \frac{8}{\tau_B}C_3 - \frac{2}{\tau_c}C_0
\ee
$V_{ij}b_{ik}v_k v_j$ :\\
\be
\frac{4}{\tau_B}C_0 + \frac{4}{\tau_c}C_3 = 0
\ee
$V_{ij}b_{ik}v_k b_j (\vec v \cdot \vec{b})$ :\\
\be
\frac{2}{\tau_B}(2C_2) + \frac{1}{\tau_c}(-4C_3 + 4C_4) = 0
\ee
$V_{ij}v_i b_j (\vec v \cdot \vec{b})$ :\\
\be
\frac{2}{\tau_B}(-8C_3 + 2C_4) - \frac{1}{\tau_c}(4C_2) = 0
\ee
Solving above four equations we finally get $C$s as,
\bea
\nn
&&C_0 = -\frac{\omega}{2T}\;
\frac{\tau_c}{\left\{1+4(\tau_c/\tau_B)^2\right\}} ~ f_0(1-a f_0)\\
\nn
&&C_2 = -\frac{\omega}{2T}\; \frac{3~ \tau_c~ (\tau_c/\tau_B)^2}{\left\{1+(\tau_c/\tau_B)^2\right\} \left\{1+4(\tau_c/\tau_B)^2\right\} } ~ f_0(1-a f_0)\\
\nn
&&C_3 = -\frac{\omega}{2T}\; \frac{\tau_c(\tau_c/\tau_B)}{\left\{1+4(\tau_c/\tau_B)^2\right\}}
~ f_0(1-a f_0)\\
&&C_4 = -\frac{\omega}{2T}\; \frac{\tau_c(\tau_c/\tau_B)}{1+(\tau_c/\tau_B)^2} ~ f_0(1-a f_0)
\label{C'sR}
\eea

\section{Appendix: massless results}
\label{App}
\subsection{Thermodynamics of massless quark at $B=0$}
\label{Ap_Th_B0}
Here we have shown explicit calculation of energy density, pressure and entropy density for quasi particle system considering mass $m=0$ for MB, BE, FD at $B=0$.

Thermal distribution function can be written in a general way in the form 
\bea
f = \frac{1}{e^{\beta E} + a} 
\eea
where, $a=0$ for MB, $a=1$ for FD and $a=-1$ for BE statistics. \\
\textbf{BE:}
Energy density for bosons is given by
\bea
\epsilon = g\int_0^{\infty}\frac{d^3p}{(2\pi)^3}\frac{E}{e^{\beta E}-1}
\eea
Here $E = p$ which gives us 
$$dp = dE$$
and the integral becomes
\bea
\epsilon &&= g\int_0^{\infty} \frac{4\pi E^3dE}{(2\pi)^3}\frac{1}{e^{\beta E} -1}\nn \\
&& = g\int_0^{\infty}\frac{E^3dE}{2\pi^2}\frac{1}{e^{\beta E} -1}
\eea
Substituting $\beta E = x$ giving 
$$dE = \frac{dx}{\beta}$$
gives us 
\bea
\epsilon = \frac{g_b}{2\pi^2\beta^4} \int_0^{\infty}\frac{x^3dx}{e^x-1} 
\eea
This integral can be converted into a $\zeta(s)$ function by using 
\bea
\zeta(s) = \frac{1}{\Gamma(s)}\int_0^{\infty}\frac{x^{s-1}dx}{e^x-1} 
\eea
where $\Gamma(s)$ is the gamma function. 
\bea
\epsilon &&= g \times \frac{(k_BT)^4}{2\pi^2}\zeta(4)\Gamma(4)\nn \\
&& = g\frac{\pi^2}{30}T^4
\eea
Using $\zeta(4) = \frac{\pi^4}{90}$ and $\Gamma(4) = 3! $
The corresponding pressure density is 
\bea
P = \frac{\epsilon}{3} = gT^4\frac{\pi^2}{90}
\eea
The entropy density is $$ s = \frac{\epsilon + P}{T} = \frac{4g\pi^2}{90}T^3$$

\textbf{FD:} Energy density for fermion is 
\bea 
\epsilon = g \int \frac{d^3p}{(2\pi)^3}E \frac{1}{e^{\beta E} + 1}
\eea
$E = p$ which gives the integral 
\bea
\epsilon = \frac{g}{2\pi^2}\int_0^{\infty} \frac{E^3 dE}{e^{\beta E} + 1}
\eea
Substituting $\beta E = x$ we get
$$dE = \frac{dx}{\beta} $$
\bea
\epsilon &&= \frac{g}{2\pi^2}T^4\Gamma(4)(1-\frac{1}{2^{4-1}})\zeta(4)\nn \\
&& = \frac{7}{8}g\frac{\pi^2}{30}T^4
\eea
Pressure density of fermions is given by 
\bea
 P = \frac{\epsilon}{3} = \frac{7}{8}g\frac{\pi^2}{90}T^4
\eea
The entropy density is given by $$s = \frac{\epsilon + P}{T} = \frac{7g\pi^2}{90}T^3$$

\textbf{MB:} For the Maxwell Boltzmann distribution the energy density is given by 
\bea 
\epsilon = g \int_0^{\infty} \frac{d^3p }{(2\pi)^3}\frac{E}{e^{\beta E}}\nn \\
\eea
Since $E = p$ the energy density calculated by changing the integral to a beta function is
\bea
\epsilon =  \frac{3gT^4}{\pi^2}
\eea
The pressure density is $ P = \frac{\epsilon}{3}$
\bea
P = \frac{g}{\pi^2}T^4
\eea
The entropy density is given by 
\bea
s = \frac{\epsilon + P }{T} = \frac{4g}{\pi^2}T^3
\eea
\subsection{Shear viscosity and electrical conductivity of massless quark at $B=0$}
\label{Ap_Shel_B0}
Here we have shown details derivation of $\eta(T)$ and $\sigma(T)$ for MB, BE, FD at $B= 0$ for massless quasi particle system.
The shear viscosity $\eta$ for bosons is given by
\bea
\eta = \frac{g\beta}{15} \int \frac{d^3p}{(2\pi)^3}\frac{p^4}{E^2}\tau_cf_0(1+f_0)\nn \\
\eea
Here $E = p$ which gives us
\bea
\eta = \frac{g\beta}{15}\frac{\tau_c}{2\pi^2}\int_0^{\infty}p^4 dpf_0(1+f_0)
\eea
The shear viscosity for fermions is given by 
\bea
\eta&& = \frac{g\beta}{15}\int \frac{4\pi p^2 dp }{(2\pi)^3}p^2\tau_cf_0(1-f_0)\nn \\
&& = \frac{g\beta}{15}\frac{\tau_c}{2\pi^2}\int_0^{\infty}p^4dpf_0(1-f_0)
\eea
The above 2 expressions are written as 
\bea
\eta = A \int_0^{\infty}p^4dpf_0(1-f_0) = A I_1
\eea
for fermions and
\bea 
\eta = A \int_0^{\infty} p^4dp f_0(1+f_0) = AI_2
\eea
for bosons. 
where the constant A is the subsitution for 
\bea
A = \frac{g\beta}{30\pi^2}\tau_c
\eea
The $I_1$ integral is evaluated as follows
\bea
I = \int_0^{\infty} p^4 dp f_0(1-f_0)
\eea
where $ f_0 = \frac{1}{e^{\beta E}+1} = \frac{1}{e^{\beta p}+1}$ is the distribution  function for fermions. 

\bea
I_1 &&= \int_0^{\infty} p^4 dp \frac{1}{e^{\beta p}+1}[1- \frac{1}{e^{\beta p}+1}]\nn \\
&& = \int_0^{\infty} p^4 dp \frac{e^{\beta p}}{(e^{\beta p}+1)^2}\nn \\
&& = -\frac{\partial }{\partial \beta}\int_0^{\infty} \frac{p^3}{e^{\beta p} +1} dp\nn \\
\eea
By using the definition of $d(s)$ and $\zeta(s)$ function we solve the above integral as 
\bea
I_1&& = -\frac{\partial }{\partial \beta}\frac{\Gamma(4)}{(\beta)^4}(1-\frac{1}{2^3})\zeta(4)\nn \\
&& = -\frac{\partial}{\partial \beta}\frac{3!}{\beta^4}\frac{7}{8}\zeta(4)\nn \\
&& = \frac{4!}{\beta^5}\zeta(4)\frac{7}{8}
\eea
Thus $$\eta_2|_{fermions} = A \frac{4!}{\beta^5}\zeta(4)\frac{7}{8}$$\\
The integral for Bosons is 
\bea
I_2 && = \int_0^{\infty}p^4dp\Big(\frac{1}{e^{\beta p} -1}\Big)\Big(1+\frac{1}{e^{\beta p}-1}\Big)\nn \\
&& = \int_0^{\infty}p^4 dp \frac{e^{\beta p}}{(e^{\beta p}-1)^2}\nn \\
&& = -\frac{\partial}{\partial \beta}\int_0^{\infty}p^3dp \frac{e^{\beta p}}{e^{\beta p}-1}\nn \\
\eea
Substituting $\beta p = x$ we get 
\bea
I_2 = -\frac{\partial}{\partial \beta}\Big(\int_0^{\infty}dx\frac{x^3}{\beta^4(e^x-1)} \Big)
\eea
Using the definition of $\zeta(s)$ the integral is calculated as 
\bea
I_2 = \frac{4}{\beta^5}\Gamma(4)\zeta(4)
\eea
$\eta|_{Bosons}$ is obtained as
\bea
\eta|_{Bosons} = A \frac{4}{\beta^5}\Gamma(4)\zeta(4)
\eea
For Maxwell-Boltzmann distribution the shear viscosity is obtained as follows

\bea
\eta = \frac{g\beta}{15}\int \frac{d^3p}{(2\pi)^3}p^2f_0 \tau_c
\eea
where $f_0 = e^{-\beta E}$ and here $E = p$. 

\bea
\eta &&= A \int_0^{\infty} p^4 dp e^{-\beta E}\nn \\
&& = \frac{A}{\beta^5}\int_0^{\infty} e^{-x}x^4 dx \nn \\
&& = \frac{A}{\beta^5}\Gamma(5)
\eea
where $A = \frac{g\beta}{30\pi^2}\tau_c$ and $\Gamma(5) = 4!$\\

\textbf{Electrical Conductivity $\sigma$} for different distributions is calculated as follows: \\

\textbf{MB:} For Maxwell-Boltzmann distribution electrical conductivity is 
\bea
\sigma&& = \frac{q^2 g \beta}{3}\int \frac{d^3p}{(2\pi)^3}\frac{p^2}{E^2}\tau_cf_0\nn \\
&& = \frac{q^2 g \beta}{3}\int \frac{4\pi p^2 dp}{(2\pi)^3}\tau_cf_0
\eea
The distribution function is $f_0 = e^{-\beta p}$ for Maxwell-Boltzmann distribution. 

\bea
\sigma &&= \frac{q^2 g \beta}{3(2\pi^2)}\int_0^{\infty} p^2 dp [\tau_c e^{-\beta p}]\nn \\
\eea
Substituting $\beta p = x$ we get 
\bea
\sigma &&= \frac{q^2 g \beta}{3(2\pi^2)}\tau_c\int_0^{\infty} \frac{1}{\beta^3} dx  x^2 e^{-x}\nn \\
&& = \frac{q^2 g \beta}{3(2\pi^2)} \frac{2!}{\beta^3}\tau_c\nn \\
&& = q^2\frac{g}{3\pi^2}\frac{\tau_c}{\beta^2}
\eea

\textbf{FD:} The electrical conductivity of fermions is given by
\bea
\sigma = q^2\frac{g\beta}{3}\int \frac{d^3p}{(2\pi)^3}\frac{p^2}{E^2}\tau_cf_0(1-f_0)
\eea
where the distribution function of fermions is given by 
$$f_0 = \frac{1}{e^{\beta p} +1}$$
\bea
\sigma&& =  q^2\frac{g\beta}{3} \int _0^{\infty} \frac{4\pi p^2 }{(2\pi)^3}dp \times \tau_c\frac{e^{\beta p}}{(e^{\beta p} +1)^2} \nn \\
&& = \frac{q^2 g \beta}{3\times 2 \pi^2}\int_0^{\infty}p^2 dp \frac{e^{\beta p}}{(e^{\beta p}+1)^2}\tau_c
\eea
Using the definition of $\zeta(s)$ the integral is evaluated to be 
\bea
\sigma = \frac{q^2 g}{2\times 2\pi^2}\frac{1}{\beta^2}\frac{\Gamma(4)}{4}\zeta(3)\tau_c
\eea
where $\Gamma(4) =3!$ \\
\textbf{BE:} For bosons $\sigma$ is 
\bea
\sigma = \frac{q^2g\beta}{3}\int \frac{d^3p}{(2\pi)^3}\frac{p^2}{E^2}f_0(1+f_0)\tau_c
\eea
where the distribution function of bosons is givn by 
$$f_0 = \frac{1}{e^{\beta p} -1}$$
\bea
\sigma &=& \frac{q^2g\beta}{3\times 2\pi^2}\int_0^{\infty} \frac{p^2 e^{\beta p}}{(e^{\beta p}-1)^2}dp \tau_c 
\nn\\
&=& \frac{q^2g\beta}{3\times 2\pi^2} \times I \times \tau_c
\eea
where the integral I is given by 
\bea
I = \int _0 ^{\infty} \frac{p^2 e^{\beta p}}{(e^{\beta p} -1)^2}dp  
\eea
is solved as follows 

\bea
I &&= -\frac{\partial }{\partial \beta}\int_0^{\infty} \frac{p}{e^{\beta p}-1}dp\nn \\
&& = -\frac{\partial}{\partial \beta}\Big(\frac{1}{\beta^2}\Big)\int_0^{\infty} \frac{x}{e^x-1}dx \nn \\
&& = \frac{2}{\beta^3}\int_0^{\infty}\frac{x}{e^x-1}dx\nn \\
&& = \frac{2}{\beta^3}\zeta(2)\Gamma(2) = \frac{2}{\beta^3}\zeta(2) 
\eea
Putting this in the expression for conductivity we get 
\bea 
\sigma = \frac{q^2 g \beta}{3\times 2 \pi^2}\frac{2}{\beta^3}\zeta(2)\tau_c 
\eea
\subsection{Thermal average Energy}
\label{Ap_Eav}
The expressions of viscosity and conductivity contain magnetic relaxation time $\tau_B$ which is $\tau_B=\frac{E}{qB}$. But for simplicity of calculation we will consider average energy for $\tau_B$ calculation. So, $\tau_B=\frac{\langle E \rangle}{qB}$\\
\textbf{MB:} Average energy for Maxwell Boltzmann distribution with $E =p$
\bea
\langle E \rangle &&= \frac{\int \frac{d^3p}{(2\pi)^3}E e^{-\beta E}}{\int\frac{d^3p}{(2\pi)^3}e^{-\beta E}}\nn \\
&& = \frac{\int_0^{\infty}p^3dpe^{-\beta p}}{\int p^2e^{-\beta p}dp}\nn \\
&& = \frac{\frac{6}{\beta^4}}{\frac{2}{\beta^3}} = \frac{3}{\beta}
\eea
\textbf{FD:} Average energy for fermions is 
\bea
\langle E \rangle &&= \frac{\int \frac{d^3p}{(2\pi)^3}E \frac{1}{e^{\beta E}+1}}{\int\frac{d^3p}{(2\pi)^3}\frac{1}{e^{\beta E}+1}}\nn \\
&& = \frac{\int_0^{\infty} \frac{p^3 dp}{e^{\beta p}+1}}{\int_0^{\infty}\frac{p^2dp}{e^{\beta p}+1}}\nn \\
\eea
Using the definition of $\zeta(s)$ and substituting $\beta p =x$ the above integral is evaluated as 
\bea
\langle E \rangle &&= \frac{\frac{\Gamma(4)}{\beta^4}\Big(1-\frac{1}{2^{4-1}}\Big)\zeta(4)}{\frac{\Gamma(3)}{\beta^3}\Big(1-\frac{1}{2^{3-1}}\Big)\zeta(3)}\nn \\
&& = \frac{7}{2}T\frac{\zeta(4)}{\zeta(3)}
\eea

\textbf{BE:} Average energy of Bosons  with $E =p$

\bea
\langle E \rangle &&= \frac{\int\frac{d^3p}{(2\pi)^3}\frac{E}{(e^{\beta E}-1)}}{\int\frac{d^3p}{(2\pi)^3}\frac{1}{(e^{\beta E}-1)}} \nn \\
&& = \frac{\int_0^{\infty}\frac{p^3 dp}{(e^{\beta p}-1)}}{\int _0^{\infty}\frac{p^2 dp}{(e^{\beta p}-1)}}
\eea
Using the definition of $\zeta (s)$ and the substitution $\beta p =x $ the integral can be solved as 
\bea
\langle E \rangle &&=  \frac{1}{\beta} \frac{\zeta(4)\Gamma(4)}{\zeta(3)\Gamma(3)}
\eea

\end{document}